\documentclass[12pt,preprint]{aastex}

\begin{document}

\title{The Hydromagnetic Interior of a Solar Quiescent Prominence. \\
I. Coupling between Force-balance and Steady Energy-transport.}

\author{B. C. Low\altaffilmark{1}, T. Berger\altaffilmark{2}, R. Casini\altaffilmark{1}, \& W. Liu\altaffilmark{2,3}\\
March 5, 2012}

\altaffiltext{1}{High Altitude Observatory, National Center for Atmospheric
Research, P.O. Box 3000, Boulder, CO 80307, USA}
\altaffiltext{2}{Lockheed-Martin Advanced Technology Center, Solar and Astrophysics Laboratory,
3251 Hanover St., Palo Alto, CA 94304, USA}
\altaffiltext{3}{W. W. Hansen Experimental Physics Laboratory, Stanford University, Stanford, CA 94305}
%%%%%%%%%%%%%%%%%%%%%%%%%%%%%%%%%%%%%%%%%%%%%%%%%%%%%%%%%%%%%
 
\begin{abstract}
\noindent
This series of papers investigates the dynamic interior of a quiescent prominence revealed by recent {\it Hinode} and {\it SDO/AIA} high-resolution observations.   This first paper is a study of the static equilibrium of the Kippenhahn-Schl\"{u}ter diffuse plasma slab, suspended vertically in a bowed magnetic field, under the frozen-in condition and subject to a theoretical thermal balance among an optically-thin radiation, heating, and field-aligned thermal conduction.  The everywhere-analytical solutions to this nonlinear problem are an extremely restricted subset of the physically admissible states of the system.  For most values of the total mass frozen into a given bowed field, force-balance  and steady energy-transport cannot both be met without a finite fraction of the total mass having collapsed into a cold sheet of zero thickness, within which the frozen-in condition must break down.  An exact, resistive hydromagnetic extension of the Kippenhahn-Schl\"{u}ter slab is also presented, resolving the mass-sheet singularity into a finite-thickness layer of steadily-falling dense fluid.  Our hydromagnetic result suggests that the narrow, vertical prominence $H_{\alpha}$ threads may be falling across magnetic fields, with optically-thick cores much denser and ionized to much lower degrees than conventionally considered.  This implication is discussed in relation to (i) the recent {\it SDO/AIA} observations of quiescent prominences that are massive and yet draining mass everywhere in their interiors, (ii) the canonical range of $5-60~G$ determined from spectral-polarimetric observations of prominence magnetic fields over the years and (iii) the need for a more realistic multi-fluid treatment.  
\end{abstract}


%%%%%%%%%%%%%%%%%%%%%%%%%%%%%%%%%%%%%%%%%%%%%%%%%%%%%%%%

\section{Introduction}

Quiescent prominences are long-lived, partially-ionized, plasma condensations in the tenuous, million-degree hot,
ionized solar corona \citep{martin94,th95,gaiz98,labrosse10,mackay10}.  These partially-ionized condensations, at least, their optically-thin 
observable parts, are typically two orders of magnitude denser and cooler than the surrounding corona.  
As a macroscopic structure prior to an eruption, a prominence can be remarkably stable, persisting for days to weeks as a long, vertical slab, approximately $5 \times 10^3 ~ km$ wide, $2 \times 10^4 ~ km$ tall, $10^5 ~ km$ long, seemingly suspended by magnetic fields to curve the full prominence length right above and along a magnetic neutral line on the photosphere.
By prominence we shall henceforth be referring to this common type, the polar crown prominence being the canonical example; see the above review articles for a comprehensive view of prominences of all varieties.   

Figure 1 is a Hinode/SOT image taken in the Ca II 396.8 nm "H-line" spectral region showing the cool part of the prominence at a high spatial resolution, revealing its characteristic narrow, vertical, dense threads interspersed among narrow, dark lanes.  This static image belies the remarkably dynamic nature of these interior structures.  Prominence movies from {\it Hinode} and {\it SDO/AIA} instruments have established that over internal scales of the order of $300~km$ and time scales of minutes, these structures are in a constant state of motion at less than free-fall speeds in a characteristic range of $5-30~ km~s^{-1}$ \citep{berger08, berger10, berger11, okamoto07,okamoto10}.    The ever-present cool, dense threads are falling whereas the dark lanes among them are narrow streams of hot, tenuous plasma traveling upward at comparable speeds  \citep{berger08, berger10}.  Vertical motions in both directions are actually present in both types of structures, but the high-resolution movies show a discernible average descent of the dense threads and a general bubbly rise of hot, tenuous plasmas.  The latter may be the result of rapid heating to coronal temperatures associated with emerging magnetic fluxes that push into the overlying dense prominence material.  This condition may be understood in terms of a nonlinear Rayleigh-Taylor hydromagnetic instability \citep{berger10,hillier11}.  Such emergent-flux structures can be of considerable sizes that first form quasi-statically beneath a quiescent prominence before pushing their ways clear through the prominence without a macroscopic eruption \citep{detoma08,berger11}.

A recent quantitative analysis of the motions of descending threads observed with {\it SDO/AIA} indicates that over a 20 hour period, an order of magnitude more mass may be drained in this manner than found at any one time in a  moderate-size, slowly evolving prominence \citep{liu12}.  Using the lower end of the range of observed prominence densities to compute, the drainage can remove some $10^{15}~g$ from the corona in a day, a mass comparable to that of a coronal mass ejection. Of particular interest is that the drainage rate estimated from observation is changing in proportion to the slowly-changing, estimated total prominence mass during quasi-steady evolution.   This observational result raises two questions.  How does the observed significant mass drainage via cool threads relate to the replenishing mass-injection represented by the upward streams of hot, tenuous plasmas?  Berger et al. (2011) propose that emerging magnetic flux drives a constant cycling of mass from the photosphere, in the form of tenuous hot plasma, into the coronal cavity overlying a quiescent prominence.  This up-flow returns as the draining vertical threads, essentially establishing a novel form of 'magneto-thermal' convection in the solar outer atmosphere.

The other question relates to the embedded magnetic fields in these processes.   Spectral polarimetric observations have shown that quiescent prominence plasmas embed typically horizontal, local fields of the order of $5-60 ~ G$ \citep{casini03,leroy89,lopez03,trujillo02}.   These fields are sufficiently intense to dictate the structures of their prominences under a high degree of the frozen-in condition due to high electrical conductivity.   In particular, a large amount of mass can be supported by fields of the observed intensities.  Yet more mass can drained through the prominence in less than a day than the amount trapped quasi-steadily in the prominence at any one time.  How is the observed drainage reconcilable with the observed prominence field being generally horizontal if we accept the frozen-in condition?  

An attractive hypothesis is that it is in the nature of prominence dynamics to develop such extreme physical conditions that the usual frozen-in condition breaks down spontaneously.   This hypothesis is what we set out to investigate as a theoretical hydromagnetic possibility.  The first paper of the series here treats the case of thermal collapse during plasma condensation producing cold, dense, poorly ionized plasmas within which the frozen-in condition cannot be sustained.  The second paper takes up a separate piece of physics, demonstrating that despite the high magnetic Reynolds-numbers of the prominence interior, magnetic reconnection takes place readily via spontaneous formation and dissipation of electric current-sheets \citep{parker94,petrie05, janse10b}.  The two papers provide a hydromagnetic basis to understand the observed restless interior of a prominence.

Specifically the first paper here treats the following theoretical problem.  We construct a complete family of exact solutions describing a vertical Kippenhahn-Schl\"{u}ter (KS) plasma-slab supported against gravity in a bowed magnetic field in unbounded space, varying only in the direction perpendicular to the slab \citep{ks57, low75, hillier11}.   We assume static force equilibrium under (i) the frozen-in condition and (ii) a steady balance among optically-thin radiation, heating, and thermal conduction directed along the field.  These three energy processes take the theoretical forms studied by Low \& Wu (1981), representing an energy transport of a minimum physical complexity relevant to prominences.  Despite its one-dimensionality, the KS slab captures important properties of this complex system.   The novel result of the study is that the set of everywhere-analytical solutions found in the earlier study is only a subset of measure zero of all the physically admissible equilibrium states of the KS slab.   Generally, force-balance and steady energy-transport cannot both be met under the frozen-in condition without a fraction of the total mass of the slab having collapsed into a mathematical singularity representing an exceedingly cold, dense sheet at the center of the otherwise diffuse slab.   The weight of that mass sheet is supported by a discrete Lorentz force due to a discrete electric current in the sheet.   This discrete current can be maintained only if the frozen-in condition is rigorously obeyed, that is, the electrical conductivity is infinite.    The electrical conductivity in the solar atmosphere is large but finite, of course.  In that case, an evolution to equilibrium must at first seek a state with an inevitable singularity when the frozen-in condition does apply to a very high degree.  As the singularity develops, the small nonzero electrical resistivity unavoidably becomes significant at a point in time, resulting in current dissipation.  A breakdown of the frozen-in condition thus occurs spontaneously.

The general inevitability of mass-sheet singularity in the KS-slab is demonstrated in Section 2.  The singular equilibrium solutions containing a mass sheet with a discrete current are analyzed in Section 3.   The breakdown of the frozen-in condition is illustrated in Section 4, treating an exact solution for a steady, vertical resistive hydromagnetic flow and pointing out the need for a more realistic multi-fluid description.  Our results are summarized in Section 5 to relate in general terms to the physics of prominences .

\section{Coupling between static force-balance and steady energy-transport}

We adopt a one-fluid hydromagnetic description of the corona and prominence:
\begin{eqnarray}
\label{momentum}
\rho\left[{\partial {\bf v} \over \partial t} + \left({\bf v} \cdot \nabla \right) {\bf v} \right] &=& {1 \over 4 \pi}(\nabla \times {\bf B} ) \times {\bf B} -\nabla p -\rho g {\hat z}  , \\
\label{induction_resistive}
{\partial {\bf B} \over \partial t} &=& \nabla \times \left( {\bf v} \times {\bf B} - \eta \nabla \times {\bf B} \right) ,
\end{eqnarray}
\noindent
coupling the velocity ${\bf v}$ to the magnetic field ${\bf B}$; $\rho$, $p$, and $g$ being the density, pressure and uniform gravitational acceleration in the negative Cartesian-$z$ direction.  The coefficient of resistivity $\eta = c^2/4 \pi \sigma$ is generally very small over all observable length scales in the sense of a large magnetic Reynolds-number $R_M=v_0l_0/\eta$, $c$ and $\sigma$ being the speed of light and electrical conductivity, respectively, in cgs units, and, $v_0$ and $l_0$ denoting characteristic speed and length, respectively.  The time $\tau_D = l_0^2/\eta$ for a magnetic structure of characteristic length $l_0$ to diffuse resistively is another indicator of the importance of resistive effects.   Set $l_0 \approx 7 \times 10^7 cm$ as the limit of spatial resolutions achievable with a ground-based telescope.  Then the Spitzer (1962) resistivity $\eta=3 \times 10^{12} T^{-3/2}~cm^2~s^{-1}$, where $T$ is the temperature, gives $\tau_D$ well in excess of tens of years or longer for temperatures in the range of $10^{4} -10^{6~o}K$.  Therefore, the frozen-in condition can only break down if a hydromagnetic fluid develops exceedingly small scale structures.  In the case of the corona, these scales may even fall below the mean free path of the fluid, so that the applicable resistivity involves coupling between the large-scale fluid and particle-kinetic effects \citep{low12}. 

We are interested in a static equilibrium with a negligible $\eta$, so that currents may persist permanently.   Then, setting ${\bf v} = 0$, the induction equation (\ref{induction_resistive}) is irrelevant except to demand that ${\bf B}$ is solenoidal.  The mass-conservation equation is also irrelevant, and static equilibrium is described by the complete set of equations:
\begin{eqnarray}
\label{fb}
{1 \over 4 \pi}(\nabla \times {\bf B} ) \times {\bf B} -\nabla p -\rho g {\hat z} = 0 , \\
\label{solenoid}
\nabla \cdot {\bf B}  = 0,  \\
\label{energy}
\nabla \cdot \left[  \kappa { \left ( {\bf B} \cdot \nabla T \right) {\bf B} \over |{\bf B}|^2 } \right]  -  r(\rho, T) + h(\rho,T) = 0 , \\
\label{ideal_gas}
p = {k_B \over m_0} \rho T ,
\end{eqnarray}
\noindent
where we introduce the ideal gas law, with $k_B$ and $m_0$ being the Boltzmann constant and the mean mass of the fluid particles, respectively.   Force-balance equation (\ref{fb}) is coupled to steady energy equation (\ref{energy}).   The latter balances radiative loss $r$, heat gain $h$ and a thermal conductive flux aligned along ${\bf B}$ with a thermal conductivity $\kappa$.   The properties of a specific plasma fluid is represented by $r$, $h$ and $\kappa$ as known functions of the thermodynamic variables, $p$, $\rho$ and temperature $T$.  

This radiative magnetostatic problem is formidable, especially for a three-dimensional system.  We therefore treat a localized one-dimensional system.  To keep physical issues in mind we first briefly describe the model in relation to the real prominence and its surrounding.  Then we formulate the mathematical problem in Section 2.1 and show in Section 2.2 that the physically complete set of solutions must include both everywhere-analytic solutions as well as the solutions with admissible singularities presented in Section 3.

The prominence in Figure 1 belongs to a larger-scale, three-part coronal helmet-streamer of the kind in the white-light image of a total eclipse in Figure 2.  Coronal observation of the magnetic fields external to prominences in the corona is still in its infancy \citep{casini99, judge06,tomczyk08, dove11}, but it is physically reasonable to interpret a helmet-streamer to be an arcade of closed magnetic field anchored with magnetic footpoints frozen into the base of the corona \citep{low96, low01,zhang05}.  This field traps the high density of the helmet-streamer in approximate static equilibrium under the frozen-in condition.  External to the helmet-streamer is the solar wind accelerated from quasi-static conditions at the coronal base.  This wind combs its frozen-in fields into an open configuration, each field line extending from one footpoint anchored in the coronal base out into the heliosphere. The bright helmet often has a tenuous, dark cavity at its base within which a quiescent prominence can be found located at the bottom of that cavity.  When the line of sight is aligned favorably with the arcade of the helmet-streamer, an exemplary case in Figure 2, this three-part structure is seen clearly.   

Observations support the interpretation that the cavity is a magnetic flux rope with field lines that wind around within that cavity. \citep{fan01,manchest04,zhang05,cottaar09}.  The prominence may then be explained as the cool condensed plasma trapped at the bottom of local U-shaped flux tubes under a high degree of the frozen-in condition \citep{fan06,fan07,fuller08,gibson06,gibson10,schmit09,schmit11}.  The frozen-in condition permits no motion of the condensation across magnetic flux surfaces.  Our study here sets out to first understand this frozen-in state well, only to discover from the physics of such a state that the frozen-in condition must generally break down in the prominence interior to produce cross-field transport of mass at less than free-fall speeds. 

\subsection{A thermally-balanced Kippenhahn-Schl\"{u}ter slab}  

Consider a 1D system that depends on only the Cartesian coordinate $y$.  Let the field have the solenoidal form:
\begin{equation}
\label{KS_B}
{\bf B} = B_0 \left[ G_0, 1, H(y) \right] ,
\end{equation}
\noindent
$B_0$ and $G_0$ being constants. We are interested in $B_z = B_0H(y)$ being a monotonically-increasing, odd function with $H = 0$ at $y = 0$.   This field is deformed from a uniform "background" field 
\begin{equation}
\label{B_bg}
{\bf B}_{bg} = B_0 \left[ G_0, 1, 0 \right] ,
\end{equation}
\noindent
by weighing upon it with frozen-in mass, producing a bow-shape geometry to represent a local U-shaped.  In this idealized unbounded domain, the surrounding prominence environment tends to a uniform state located at $y \rightarrow \pm \infty$.  

Introduce the dimensionless variables, 
\begin{equation}
\label{norm}
P = p/p_0 , ~~D = \rho/\rho_0, ~~ \theta = T/T_0, ~~ Y = {y \over L_0} ,
\end{equation}
\noindent
where $L_0 =  {k_BT_0 \over m_0 g}$ is the hydrostatic density scale height at temperature $T_0$ related to the other normalization constants by $p_0 = \rho_0 kT_0/m_0$.   The ideal gas law is then 
\begin{equation}
\label{ideal_gas2}
P = D \theta .
\end{equation}
\noindent
Introduce the two plasma $\beta$-parameters:
\begin{eqnarray}
\label{beta_0}
\beta & = & {8 \pi p_0 \over B_0^2} , \\
\label{beta}
\beta_{net} & = & {8 \pi p_0 \over B_0^2\left( 1 + H(Y)^2 + G_0^2 \right)}  < \beta ,
\end{eqnarray}
\noindent
respectively measuring the importance of the fluid pressure relative to the magnetic pressures of the uniform $B_y = B_0$
and the total field $|{\bf B}|=B_0 \left( 1 + H^2 + G_0^2 \right)^{1/2}$.  With no loss of generality, take $p = p_0$ to be the pressure at the center of the sheet, ie, $P = 1$, $H = 0$ at $Y = 0$.  Then, the force-balance equations (\ref{fb}) reduces to the $y$ and $z$ components:  
\begin{eqnarray}
\label{pressure}
\beta P + H^2 = \beta , \\
\label{density}
{dH \over dY} = {1 \over 2} \beta D . 
\end{eqnarray}
\noindent
Equation (\ref{pressure}) shows that the total pressure $p + B^2/8 \pi$ is uniform in space which is a sufficient condition for linear hydromagnetic stability \citep{zweibel82}.   The high pressure of the sheet is confined by the vertical component $B_z = B_0 H$ significant outside the sheet.   
Equation (\ref{fb}) can be written in the form
\begin{equation}
\label{fb2}
{1 \over 4 \pi} \left( {\bf B} \cdot \nabla \right) {\bf B} -\nabla \left( p + {B^2 \over 8 \pi} \right) -\rho g {\hat z} = 0 , 
\end{equation}
\noindent
in terms of the magnetic tension and the total-pressure force.  Since the total pressure of the KS-slab is uniform in space, the tension force is everywhere vertical and it alone supports the weight of the plasma.  In other words, the left hand side of Equation (\ref{density}) is essentially the upward magnetic tension force. 

We pause to relate the above idealized magnetic field to spectral-polarimetric observations of prominences.  If we identify the concentration of plasma around $Y=0$ with the prominence plasma being observed, where $B_z=B_0H \approx 0$, polarimetric observation will detect the horizontal field ${\bf B}_{bg}$ given by Equation (\ref{B_bg}).   The observed field has intensity $B_0\left(1 + G_0^2\right)^{1/2}$ which can fall within the observed range of $5-60~G$ even with $B_0$ small as long as $G_0$ is sufficiently large.  Therefore, the plasma-beta $\beta$ given by Equation (\ref{beta_0}) may be larger than unity whereas the plasma-beta $\beta_{net}$ defined by Equation (\ref{beta}) in terms of the net field is typically less than unity.  This feature simply reflects the fact that a slab of plasma hemmed up by a background field aligned close to the central plane of the slab will snag only a weak component $B_y$ of the field across it.  Therefore, the field would sag greatly in the $y-z$ plane with a large value of $\beta$.

We take advantage of this simplicity of the model to deal with the complexity of thermal balance.  We adopt the temperature-dependent Spitzer (1962) thermal conductivity
\begin{equation}
\label{spitzer}
\kappa = \kappa_S~T^{5/2}~erg~cm^{-1}s^{-1}K^{-1} ,
\end{equation}
\noindent
for a fully ionized gas where $\kappa_S = 2 \times 10^{6}$.  For our theoretical purpose, we adopt this conduction model in a formal sense with $\kappa_S$ taken as an ad hoc constant, so as to be conceptually specific about the model of thermal conduction adopted.  To complete the specification of the fluid's material properties, we take
\begin{eqnarray}
\label{radiation}
r & = & \alpha_0 \rho^2 T , \\
\label{heating}
h & = & \gamma_0 \rho ,
\end{eqnarray}
\noindent
where $\alpha_0$ and $\gamma_0$ are constants characteristic of the fluid.  Here we model radiative loss $r$ to be optically thin, collisional in nature as represented by its quadratic dependence on density.  Detailed radiative calculations  have shown that the collisional $r$ has this form with $\alpha_0$ approximately constant in contiguous ranges of temperatures from $10^4 K$ to $10^6K$, multiplied by a monomial in $T$ of a constant power in each of those ranges \citep{th95}.  In Equation (\ref{radiation}) we take $\alpha_0$ to be constant and the temperature factor to be linear in $T$ to simplify an otherwise very complex mathematical problem.  The origin of heating in the prominence is poorly understood, representing another uncertainty in such thermal models.  For our theoretical study we use an ad hoc heating rate proportional to density \citep{uchida63,th95}.   These choices lead to an analytically integrable model \citep{low81}.  We wish to investigate this model afresh to understand the physical meaning of these analytical solutions.

With the material properties of the plasma fluid so specified, energy equation (\ref{energy}) takes the dimensionless form:
\begin{equation}
\label{energy2}
{d \over dY} \left[ {\kappa_0 \theta^{5/2} \over 1 + G_0^2 + H^2} {d\theta \over dY} \right] = D^2 \theta - \gamma D ,
\end{equation}
\noindent
where 
\begin{eqnarray}
\label{gamma}
\gamma & =& {\gamma_0 k_B\over \alpha_0 m_0 p_0} =  {8 \pi \gamma_0 k_B\over \alpha_0 m_0 B_0^2} {1 \over \beta} , \\
\label{kappa_0}
\kappa_0 & = & {\kappa_S T_0^{5/2} \over \alpha_0 \rho_0^2L_0^2} ,
\end{eqnarray}
\noindent
$L_0$ being the scale height at temperature $T_0$ already introduced.   Equation (\ref{energy2}) as a second order ODE requires two boundary conditions to determine $\theta$.   We choose them to be
\begin{eqnarray}
\label{theta_bc1}
&Y& = 0  ,~~~ {d \theta^{7/2} \over dY} = 0 , \\
\label{theta_bc2}
&Y& \rightarrow \infty, ~~~ {d \theta^{7/2} \over dY} \rightarrow 0 .
\end{eqnarray}
\noindent
Condition (\ref{theta_bc1}) comes from the assumed symmetry of $\theta(Y)$ about $Y=0$.  Since the heating source is explicitly accounted for by $h$, condition (\ref{theta_bc2}) sets the thermal flux to vanish at infinity.  In the unbounded model considered here, the $h$-heating and  $r$-radiation vanish at infinity.  An external source of thermal flux coming in from infinity would require an unphysical temperature that increases to unbounded values at infinity.  An external sink of thermal flux at infinity is incompatible with the unbounded domain, since a monotonically decreasing temperature must become zero at some finite $Y$ on either sides of $Y = 0$.  The $h$-heating is the only source of energy in this magnetostatic atmosphere with thermal conduction transporting excess heat to where it can be radiated away.   The equations  (\ref{ideal_gas2}), (\ref{pressure}), (\ref{density}), and (\ref{energy2}) are a complete set of scalar equations for the dependent variable $P$, $D$, $\theta$, and $H$.   In principle, Equations (\ref{ideal_gas2}), (\ref{pressure}), and (\ref{density}) determine $P$, $D$ and $H$ in terms of $\theta$ so that Equation (\ref{energy2}) determines $\theta$ as a function of $Y$ subject to the two homogeneous boundary conditions (\ref{theta_bc1}) and (\ref{theta_bc2}).  From the solution $\theta(Y)$ obtained, the other dependent variables can then be expressed as functions of $Y$.

The model is 1D only in the mathematical sense.  The field is sheared and bowed in 3D space.  This 3D geometry strongly influences energy-balance, represented by the factor $(1 + G_0^2 + H^2)^{-1}$ in Equation (\ref{energy2}).  The thermal flux is completely aligned along the field.  This factor depends on the magnitudes of the two field components, $B_x = B_0 G_0$ and $B_z = B_0 H$.  For large values of $G_0$ and $H$, respectively describing high magnetic shear and great sag in the field, the thermal flux enters the plasma slab at a large oblique angle.  

\subsection{The physically complete set of solutions}

The governing Equations (\ref{ideal_gas2}), (\ref{pressure}) and (\ref{density}) and (\ref{energy2}) can be reduced to a quadrature.  This was shown by Low \& Wu (1981) using $P$ as an independent variable and $D$, $\theta$, $H$ and $Y$ the dependent variables.  In that formulation, $D$, $\theta$, $H$ can be solved analytically explicitly in terms of $P$, following which solving $Y$ as a function of $P$ gives the distributions of the equilibrium field and plasma in space.  Here we take a different approach, choosing $H$ instead of $P$ as the independent variable.    

By Equation (\ref{density}) describing the vertical balance of force in a slab symmetrical about $Y=0$, $H(Y)$ must be monotonically increasing in the domain $-\infty<Y<\infty$ since density $\rho=\rho_0D$ is positive.   Define $H(\pm \infty)=\pm H_0$, and Equation (\ref{pressure}) gives $H_0=\sqrt{\beta}$ with $P(\pm \infty)=0$.   The total mass $M$ in an infinitely long column, of unit cross-sectional area, parallel to the $Y$-axis is given by  
\begin{eqnarray}
\label{M0}
M g & = & \rho_0 L_0 g \int_{-\infty}^{\infty} D dY  \nonumber \\
       & = & {2 p_0 \over \beta} \int_{-\sqrt{\beta}}^{\sqrt{\beta}} dH  \nonumber \\
       & = &  {B_0^2 \over2 \pi} \sqrt{\beta} .
\end{eqnarray}
\noindent 
Hereafter we called $M$ simply the total mass.   During an evolution of this 1D system, $M$ is a constant in time.  When this system is in equilibrium in a background field with a given $B_0$, $M$ determines $\beta=8\pi p_0/B_0^2$ by Equation (\ref{M0}).  In other words, $M$ determines the central fluid pressure $p(0)=p_0$ and the far vertical field component $B_z(\pm \infty)=\pm B_0H_0$, recalling $H_0=\sqrt{\beta}$.  The greater the total mass the greater is the sag of the field with a large far vertical component and a high pressure $p_0$ at the bottom of the V-shaped field.   

We now pose the following mathematical problem.  Equation (\ref{pressure}) gives $P(H)$ explicitly.   If $\theta(H)$ is also known, $D(H)$ is given by the ideal gas law (\ref{ideal_gas2}).  Then solving Equation (\ref{density}) for $H(Y)$ not only determines ${\bf B}(Y)$ but also gives $P(Y)$, $\theta(Y)$, and $D(Y)$.   Therefore, the problem posed starts with determining $\theta(H)$ in the domain $|H| \le H_0 = \sqrt{\beta}$ for a specified $\beta$.  In other words, we need to transform energy equation (\ref{energy2}) as a differential equation for $\theta$ from domain $|Y| < \infty$ to an equation in the domain $|H| \le H_0 = \sqrt{\beta}$.   Let us carry out such a transformation and solve for $\theta(H)$ in the three-step development below, leading to a family of physically complete solutions, the meaning of "physically complete" becoming clear at the end of the construction.

\subsubsection{The isothermal KS slab and related simple models}

It is instructive to consider the isothermal solutions, treated by Kippenhahn and Schl\"{u}ter (1957).  Such a state arises if the fluid does not radiate, $\alpha_0=0$, and is not heated, $\gamma_0=0$, but is thermally conducting $\kappa_S \ne 0$.  Then with $\theta = \theta_1$, a constant, and $P = \theta_1D$, Equation (\ref{density}) can be integrated to give the isothermal solution:  
\begin{eqnarray}
\label{isothermal_H}
H & = & \sqrt{\beta} \tanh {\sqrt{\beta} \over 2 \theta_1} Y , \\
\label{isothermal_density}
D & = & {1 \over \theta_1} {\rm sech}^2  {\sqrt{\beta} \over 2 \theta_1} Y  .
\end{eqnarray}
\noindent
The field lines in 3D space has uniform $B_x = B_0 G_0$ and each projects into the bowed curve in the $Y-Z$ plane:
\begin{equation}
\label{isothermal_FL}
Z - Z_0 = 2 \theta_1 \log \cosh {\sqrt{\beta} \over 2  \theta_1} Y  ,
\end{equation}
\noindent  
identified by its crossing the axis $Y = 0$ at $Z = Z_0$ in normalized unit $L_0$.   This isothermal solution is displayed in Figure 3.

Along the projected field line, the invariant frozen-in 
mass $M$ is distributed hydrostatically, decreasing with $Z$ exponentially with a scale height $L_0 \theta_1$; recall that $\theta_1$ is temperature in units of the free normalization constant $T_0$.  This exponential decrease with height combined with the self-consistent shape of the field line renders the pressure $P$ a strict function of $Y$, seen in the following two equivalent expressions,     
\begin{eqnarray}
\label{isothermal_pressure}
P & = &  {\rm sech}^2  {\sqrt{\beta} \over 2 \theta_1} Y  \nonumber \\
   & = &  \exp \left({Z_0 - Z \over \theta_1} \right) ,
\end{eqnarray}
\noindent  
the latter giving $P$ varying with height $Z$ along a fixed field line identified by the constant value $Z_0$.   As a function of $Y$, $P$ has a width proportional to the scale height $L_0\theta_1$ and inversely proportional to the total mass $M$, with $\sqrt{\beta}$ related to $M$ by Equation (\ref{M0}).  With the background field ${\bf B}_{bg}$ fixed, the greater $M$ is, the greater would be the sag in the field, and, hence, the narrower would be the peaks of $P$ and its isothermal density $D$.

On a scale large compared to the pressure width, we only see the two uniform-field parts of the bowed field lines on two sides of $Y =0$.   The mass concentration centered at $Y = 0$ appears as a sheet of zero thickness on such a scale.  On the other hand, letting $\theta_1$ go to zero decreases the scale height $L_0 \theta_1$ and the width of the mass concentration both to zero in absolute measure of length.   This happens, for example, in a fluid that is not heated but constantly radiating away energy, with a very high thermal conductivity to keep the temporally decreasing $\theta_1$ uniform in space.  If we fix $\beta$ in this process, the total mass $M$ is unchanged.  In the limit $\theta_1 \rightarrow 0$, $D$ becomes proportional to $\delta(Y)$, the Dirac delta function accounting for the unchanging mass $M$.  At the same time, $H$ given by Equation (\ref{isothermal_H}) tends to a step function jumping from $-\sqrt{\beta}$ to $\sqrt{\beta}$
upon crossing $Y = 0$ from the left.  The pressure changes such that $P = D\theta_1 \rightarrow 1$ at $Y = 0$ for all time but vanishingly
small for all points $Y \ne 0$, self-consistently describing a cold thin sheet of mass in pressure equilibrium with the external magnetic field.  In these considerations physics dictates that fluid and magnetic pressures must always be bounded whereas density may grow unboundedly so long as its integral gives a bounded mass.  This narrowing of the density peak with a fixed total mass combined with a vanishingly small temperature illustrates  a form of thermal collapse relevant to Section 2.2.3.  

Two related properties are worthy of mention when the temperature varies in space.  Equilibrium requires $B_z = B_0H(Y)$ to be monotonically increasing in $-\infty < Y < \infty$.  This ensures the pressure is positive and monotonically decreasing with $Y$.  It also ensures the density is positive but does not require the density to be a monotonically decreasing function of $Y$.   Depending on the variation of the temperature with $Y$,  the density may have two maxima located off the center $Y=0$, rather like the density inversion layer in a vertical atmosphere \citep{lerche80}.  The other property is that in the unbounded space, energy balance may be such that the pressure $P$ declines with $Y$ to become zero at some finite distance $Y = \pm Y_0$ external to which $P = D = 0$.  The mass sheet has condensed and contracted to a finite width rather like a vertical atmosphere with a finite top above which is vacuum \citep{lerche80}.   
 
\subsubsection{The everywhere-analytical solutions of Low \& Wu}

Let us transform energy equation (\ref{energy2}) in $Y$-space  into its equivalent,
\begin{equation}
\label{energy_H}
{1 \over 10} \kappa_0 \beta^2 {d \over dH}\left[ {\beta - H^2  \over 1 + G_0^2 + H^2} {d\theta^{5/2} \over dH} \right] = \beta(1 - \gamma) - H^2 ,
\end{equation}
\noindent
in $H$-space,  using Equations (\ref{ideal_gas2}), (\ref{pressure}) and (\ref{density}) to remove explicit dependance on $Y$.  The term within the square brackets on the left side of Equation (\ref{energy_H}) is essentially the thermal flux along the field.  Introduce the dimensionless thermal flux ${\mathcal F}$ defined by
\begin{equation}
\label{thermal_flux}
{\mathcal F}  = {1 \over 2} \beta^2 \kappa_0{\theta^{5/2} \over 1 + G_0^2 + H^2} {d\theta \over dY} .
\end{equation}
\noindent
Removing the explicit $Y$-dependence from its right hand side, we rewrite this definition as
\begin{equation}
\label{thermal_flux_H}
{\mathcal F} = {1 \over 10} \beta^2 \kappa_0{\beta - H^2 \over 1 + G_0^2 + H^2} {d\theta^{5/2} \over dH} ,
\end{equation}
\noindent
so that Equation (\ref{energy_H}) takes the form
\begin{equation}
\label{energy_H2}
{d {\mathcal F} \over dH} = \beta(1 - \gamma) - H^2 ,
\end{equation}
\noindent
with the integral
\begin{equation}
\label{thermal_flux_F}
{\mathcal F} = {1 \over 3} H \left[ 3 \beta (1 - \gamma) - H^2\right] + H_1 ,
\end{equation}
\noindent 
$H_1$ being an integration constant.  A second integration using Equation (\ref{thermal_flux_H}) then gives
\begin{eqnarray}
\label{theta_general}
&&{3 \over 5} \beta^2  \kappa_0\left( \theta^{5/2} - \theta_2^{5/2} \right) = \nonumber \\
&& ~~~~~~~~~~6 H_1 \left( \sqrt{\beta}-H \right) + \left[ (2-3\gamma)\beta - \Gamma_0 \right] (\beta -H^2) + {1 \over 2}(\beta -H^2)^2 \nonumber \\
&& ~~~~~~~~~~-\Gamma_0 \left[3{H_1 \over  \sqrt{\beta}} + (2-3\gamma) \beta \right] \log \left(\sqrt{\beta}-H \right) \nonumber \\
&&~~~~~~~~~~+ \Gamma_0 \left[3{H_1 \over  \sqrt{\beta}} - (2-3\gamma) \beta \right] \log \left(\sqrt{\beta}+H \right) , \nonumber \\
\end{eqnarray}
\noindent  
where $\Gamma_0=1+G_0^2+\beta$ and $\theta_2$ is an integration constant.   

The homogenous boundary conditions (\ref{theta_bc1}) and (\ref{theta_bc2}) require ${\mathcal F} = 0$ at $H=0$ and $H=H_0=\sqrt{\beta}$  in the $H$-domain, leading to $H_1=0$ and $\gamma=2/3$ and the solution
\begin{equation}
\label{theta_LW}
\theta^{5/2} = \theta_2^{5/2}  - {5 \over 3 \beta^2  \kappa_0}\left( 1+G_0^2+{1 \over 2} \beta+H^2 \right) (\beta -H^2) ,
\end{equation}
\noindent  
the everywhere analytical solution of Low \& Wu (1981).  The free integration constant $\theta_2$ is the temperature at $H=\pm \sqrt{\beta}$ corresponding to $Y \rightarrow \pm \infty$.  The temperature decreases inward monotonically to the temperature $\theta_1$ at $Y=H=0$ given by 
\begin{equation}
\label{theta0}
\theta_1^{5/2} = \theta_2^{5/2}  - {5 \over 3 \beta^2  \kappa_0}\left( 1+G_0^2+{1 \over 2} \beta \right)  .
\end{equation}
\noindent  
The integration constant $\theta_2$ is arbitrary but is physically required to be positive and bounded above such that $\theta_1 \ge 0$.  Let us replace $\theta_2$ with $\theta_1$ as the arbitrary integration constant, with no loss of generality.

Our boundary value problem with its homogeneous boundary conditions is a nonlinear eigenvalue problem that admits only one value, $2/3$, for the eigenvalue $\gamma$.  We use the notation $\gamma_c=2/3$ whenever we mean $2/3$ as a special number in this sense.  The eigenfunction $\theta^{5/2}$ corresponding to $\gamma=\gamma_c$ is highly degenerate, that is, there is an infinity of eigenfunctions, corresponding to the second free integration constant $\theta_1^{5/2}$ not fixed by the boundary conditions.  For each member of this infinite set of solutions in the $H$-domain, defining the temperature $\theta$, the governing Equations (\ref{ideal_gas2}), (\ref{pressure}) and (\ref{density}) can be used to construct the corresponding spatial distribution of field and fluid.  

The eigenvalue $\gamma=\gamma_c$ is set by the global balance of energy.  Homogenous boundary conditions (\ref{theta_bc1}) and (\ref{theta_bc2}) describe a closed thermally-conducting system.  As we have pointed out, thermal conduction is not a source of energy in this unbounded domain, serving only to conduct heat to balance the total heat input ${\mathcal H}$ against the total radiational loss ${\mathcal R}$.   These two quantities can be explicitly evaluated in the $H$-domain:
\begin{eqnarray}
\label{heat_input}
{\mathcal H} & = & \gamma \int_{-\infty}^{\infty} D dY \nonumber \\
	              & = & {4 \gamma \over \sqrt{\beta} } , \\
\label{rad_loss}
{\mathcal R} & = & \int_{-\infty}^{\infty} D^2 \theta dY \nonumber \\
	              & = & {8  \over 3  \sqrt{\beta}} ,
\end{eqnarray}
\noindent
with the ratio 
\begin{equation}
\label{gamma_2/3}
{ {\mathcal H} \over  {\mathcal R} } = {2 \over 3} \gamma .
\end{equation}
\noindent
The eigenvalue $\gamma = \gamma_c = 2/3$ follows from ${\mathcal H}={\mathcal R}$ for global energy balance, described by a whole family of everywhere-analytical solutions distinguished by the temperature $\theta_1$ at $Y=0$.

Figures 4-6 show representative solutions for different values of $\beta$ and $G_0$, each with a 
temperature profile rising from some central temperature $\theta=\theta_1$ at $Y=0$ to a hundred times higher temperature $\theta=\theta_2$ at $Y \rightarrow \infty$.  This is rendered by picking a suitable value for the constant $\kappa_0$ characterizing the importance of the thermal conductivity to the radiative loss.  The constants $(\beta, G_0)$ fix the amount of frozen-in mass $M$ and the obliqueness of the field-aligned, thermal-conductive flux supplying energy to the central part of the slab.  

Figure 4 shows a solution with $\beta = 1$ and $G_0 = 0$ for a field lying entirely in the $Y-Z$ plane, a useful reference state for comparison with the equilibria in Figures 5 and 6.  Stratification by gravity implies that radiative loss and heating, respectively, dominate in the dense region around $Y = 0$ and in the far rarefied regions.  The steady state is naturally one of high temperatures in these far regions to drive a pair of thermal fluxes inward to be radiated away.  The narrowly peaked density at $Y = 0$ is produced by this global balance of energy.  The peaked density implies an enhanced radiative loss and a corresponding depletion of the incoming thermal-conductive fluxes.  The temperature gradient is extremely steep near $Y=0$, due in part to the power-law dependence of the thermal conductivity on temperature.   

Figure 5 may be interpreted as a solution with the same $B_0$ with $G_0 = 0$ as in Figure 4, but with a greater amount of frozen-in mass $M$ so that $\beta = 5.0$.   The more pronounced sag of the more heavily loaded field is obvious from the projected field line described by $Z(Y)$.  The greater sag also implies a steeper rise in temperature as we leave $Y = 0$ owing to the more oblique incidence of the thermal-conductive flux along the field.  Alternatively, the strong $B_z = B_0H(Y)$ in this case may be said to be partial insulating the central part of the slab from the extremely hot exterior. 

The temperature gradient in the vicinity of $Y = 0$ can be further enhanced by increasing the oblique angle of incidence of the thermal flux with $G_0 \ne 0$, illustrated by the solution in Figure 6.  We recall that $\beta$ is the central plasma beta defined relative to the magnetic pressure due to the uniform component $B_y = B_0$.   If we keep the background field intensity $|{\bf B}_{bg}| = B_0( 1 + G_0^2 )^{1/2}$ fixed, we may increase $G_0$ as $B_0$ decreases and $\beta$ increases correspondingly.  One way to think of this is to take the slab in Figure 4 of a fixed mass and juxtapose it to align the plane of the slab more with the background field of a fixed intensity.   The solution in Figure 6 shows the case with the slab oriented such that $B_x/B_y = G_0 = 5$, giving $\beta = 130$.  The same slab mass as in Figure 4 is now supported in the deep sag in the field line projected on the $Y-Z$ plane, as described by $Z(Y)$.  The increased insulation effect of increased $tan^{-1} H$ and $tan^{-1} G_0$ gives rise to the extremely steep temperature gradients in the vicinity of $Y = 0$ required for energy balance.

\subsubsection{The physically complete set of magnetostatic solutions}

The set of everywhere-analytical solutions are an extremely restricted subset of the physically-realizable equilibrium states of our thermally-balanced KS slab, a point not previously known.  Among the free constant parameters of this set of solutions, $\gamma_0$, $\alpha_0$ and $\kappa_S$ define the fixed basic thermal properties of a specific fluid.  Another set, $B_0$ and $G_0$, fixes the background field.   The fluid is a perfect electrical conductor, so its total mass $M$ frozen into the field is a constant in time.  Clearly, $M$ can be prescribed arbitrarily for a given  set $(\gamma_0, \alpha_0, \kappa_S, B_0, G_0)$ that specifies the nature of the fluid and the intensity of the background field.  If the desired equilibrium state is an everywhere-analytical solution, we require $\gamma = \gamma_c = 2/3$ which, by Equations (\ref{gamma}), determines $\beta=\beta_c$ where
\begin{equation}
\label{beta_c_gamma_2/3}
\beta_c = {12 \pi \gamma_0 k_B\over \alpha_0 m B_0^2} ,
\end{equation}
\noindent 
completely fixed by the given set $(\gamma_0, \alpha_0, \kappa_S, B_0, G_0)$.  By Equation (\ref{M0}), $\beta=\beta_c$ also defines a unique equilibrium total mass 
\begin{equation}
\label{M_c}
M_c = {B_0^2 \over2 \pi} \sqrt{\beta_c} ,
\end{equation}
\noindent 
fixed independent of the above arbitrarily prescribed $M$.  Hence, unless the prescribed $M$ happens to have the value 
$M_c$, the prescribed fluid cannot be in an everywhere-analytical equilibrium.  

To summarize, if we prescribe $M=M_c$, $M_c$ fix by the given set $(\gamma_0, \alpha_0, \kappa_S, B_0, G_0)$, the fluid in equilibrium has an infinity of everywhere-analytical states available, of different $Y$ dependences, depending on the central temperature $\theta(0) = \theta_1$ as a free parameter.   Each of these equilibrium states is the end state of an evolution from some prescribed initial state, governed by the time-dependent extensions of the static equations (\ref{pressure}), (\ref{density}) and (\ref{energy2}).  The total mass $M=M_c$ is a constant in that evolution.  The central temperature $\theta_1$ defines a minimum amount of internal energy retained for each piece of the equilibrium plasma.  The $\theta_1 = 0$ equilibrium state is the lowest energy state, a preferred end state in the sense that any higher-energy $\theta_1 \ne 0$ equilibrium state may evolve in response to perturbations by losing energy.  If $\theta_1 \ne 0$, the equilibrium $D$ is finite, and $P = D\theta_1=1$ at $Y =0$.  As $\theta_1 \rightarrow 0$ parametrically, $D(Y)$ as a function of $Y$ becomes unbounded at $Y=0$, diverging as $D \approx Y^{-\nu}$ with a constant power $0  < \nu <1$.  In that limit, $D(Y)$ not a Dirac delta function in the neighborhood of $Y=0$.   That is to say, the integral of $D(Y)$ over a small interval $-Y_0 < Y < Y_0$ goes to zero as $Y_0 \rightarrow 0$.  

The interesting new point is that, for {\it all other} prescribed $M \ne M_c$ with $\beta \ne \beta_c$, a contradiction arises if we demand that the equilibrium state is everywhere analytic.  Neither the boundary conditions nor the freedom to arbitrarily prescribe $M$ can be compromised.  The physics 
of the governing ordinary differential equations (ODEs) are also not in question.  Physics avoids the implied contradiction by admitting discontinuous solutions, known as weak solutions to ordinary and partial differential equations \citep{courant62}.   The continuous parts of a weak solution are described by the governing ODEs, whereas the same physics governs the discontinuities by the integral version of these ODEs.   The equilibrium states of a fluid specified by a given set $(\gamma_0, \alpha_0, \kappa_S, B_0, G_0)$ are thus generally singular. Using set-theory language, the infinite set of continuous equilibrium states with $M=M_c$ is a subset of measure zero of all the physically admissible states.

\section{Equilibrium with a cold mass sheet}

The inviscid, compressible, supersonic flow past a fixed obstacle is a familiar example of inevitable  singularities in solutions to partial differential equations.  The flow must be sub-sonic in the vicinity of the obstacle to enable the fluid to self-communicate and navigate steadily around the obstacle.   A shock front must generally be present where the supersonic flow changes discontinuously to a subsonic flow governed by the integral version of the ideal hydrodynamic equations, the Rankine-Hugoniot equations.  This surface of discontinuity can be resolved into a continuous layer of finite thickness if we introduce additional physics significant only in that layer.  For example, the introduction of a weak viscosity resolves the shock front into a layer of a thickness determined by the strength of the viscosity.  The singular KS-slab solutions presented in this Section are analogous.   The everywhere-analytical continuous solutions, requiring a single fixed total mass $M = M_c$, are physically just too restricted.   We first survey the singular solutions to understand their physical meaning.  In the next Section, we address the additional physics one might consider to account for the internal structure of the singularity.  

\subsection{The complete absence of equilibrium}

Before proceeding, we should point out the case of complete absence of equilibrium for a KS-slab.  If radiation is absent, in the limit of $\gamma \rightarrow \infty$ in Equation (\ref{energy2}), there is no steady state because the unbalanced heat input must everywhere increase the temperature in time.   In the opposite extreme, if heating is absent, with $\gamma  \rightarrow 0$, there is only one singular steady state.  The fluid simply radiates away what heat it happens to possess to reach zero temperature, collapsing its entire mass $M$ into a sheet of zero thickness; see the isothermal solutions in Section 2.2.1.  

When both radiation and heating are present, the total mass $M$ critically determines whether energy balance is possible.  In equilibrium, the hydrostatic pressure $P(Y)$ supporting the fluid weight along ${\bf B}$ must be monotonically decreasing outward from the center of the KS slab.  With the fixed heating of the form in Equation (\ref{heating}), this stratification tends to produce a dominance of radiation around $Y=0$ and a dominance of heating in the far regions.  Therefore stratification must be consistent with a temperature $\theta(Y)$ monotonically increasing with distance away from $Y=0$, such that its inward thermal conduction brings the excess heat of the far regions to be radiated away by the dense slab center.  

Too little mass results in poor radiational loss, and no equilibrium is possible as the unbalanced heat-input keeps the temperature everywhere rising monotonically in time.  Too much mass results in the opposite effect, the inability of heating to keep up with the radiational loss due to a high-density massive slab center.  We have two possibilities in this case.  The entire mass may collapse into single massive sheet in a time-dependent evolution, never attaining a steady state and leaving vacuum outside the sheet.  The other possibility is that this collapse takes place but does not involve the entire mass.  The collapse takes out just a fraction of the total mass so that the rest of the fluid in the outer layer achieves a spatially-extended steady state.  These general expectations are met in the quantitative analysis of our KS slab given below.

\subsection{The discontinuous KS-slab solutions}

Suppose the KS slab contains a central mass sheet with a total mass $m<M$.  The monotonically increasing 
$B_z=B_0H(Y)$ jumps from a value $-H_2$ to $H_2$ upon crossing $Y=0$ from negative to positive $Y$.  This jump defines a discrete sheet current in the $x$ direction producing a Lorentz force that supports the weight of the sheet,
\begin{equation}
\label{m}
m g = {B_0^2 \over 2 \pi} H_2 , 
\end{equation}
\noindent
similar to formula (\ref{M0}).  The density distribution $D(Y)$ is a Dirac delta function in the vicinity of $Y=0$ that accounts for $m$.  In contrast the pressure $P(Y)$ must be everywhere finite, given by Equation (\ref{pressure}) to ensure pressure balance with the magnetic field, with $P(0)=1$ at $Y=0$.  The transformation $-\infty<Y<\infty ~ \rightarrow ~ -\sqrt{\beta}<H<\sqrt{\beta}$ is thus discontinuous, with the single point $Y=0$ mapped into the range $-H_2<H<H_2$, and $0<Y^2$ mapped into ${\beta}<H^2$.  

By converting an integration with respect to $Y$ into its equivalent integration with respect to $H$, we need to allow for the presence of a delta function where one occurs.   The formula (\ref{heat_input}) for the total heat input ${\mathcal H}$ remains valid in the presence of the sheet of mass $m$.   The total radiative loss needs to be re-calculated, 
\begin{eqnarray}
\label{L_H2}
{\mathcal R} & = & \int_{-\infty}^{\infty} PD dY \nonumber \\
	              & = &  {2 \over \beta} \int_{-\sqrt{\beta}}^{\sqrt{\beta}} PdH \nonumber \\
	              & = &  {2 \over \beta} \left( 2 \int_{H_2}^{\sqrt{\beta}} PdH 
	                          + \int_{-H_2}^{H_2} PdH \right)  \nonumber \\
	              & = &  {2 \over \beta^2} \left( 2 \int_{H_2}^{\sqrt{\beta}} (\beta - H^2) dH 
	                          + \int_{-H_2}^{H_2} \beta dH \right) \nonumber \\
	              & = &  {4 \over 3 \beta^2} \left( 2 \beta^{3/2} + H_2^3 \right) .	                          
\end{eqnarray}
\noindent
Global energy balance ${\mathcal R}={\mathcal H}$ is then expressed by
\begin{equation}
\label{H_2}
H_2 = (3\gamma -2)^{1/3} \sqrt{\beta} ,
\end{equation}
\noindent 
relating $\gamma$ to $H_2$, or, $m$ given by Equation (\ref{m}).  Let us fix the set $(\gamma_0, \alpha_0, \kappa_S, B_0, G_0)$ to specify a fluid of interest.  Setting the eigenvalue condition $\gamma=\gamma_c$ yields $H_2=m=0$ and $M=M_c$, recovering the everywhere-analytic solutions.  Conversely,  if $M \ne M_c$, use Equation (\ref{M0}) to fix $\beta$ in terms of the given $M$, which goes on to determine a $\gamma \ne \gamma_c$ by Equation (\ref{gamma}) and a consistent $H_2$ by Equation (\ref{H_2}).  The mapping $-\infty<Y<\infty ~ \rightarrow ~ -\sqrt{\beta}<H<\sqrt{\beta}$ requires $H_2$ to be positive and bounded above by $\sqrt{\beta}$.  Whether our $M \ne M_c$ construction yields a physical solution depends on whether $H_2$ meets the aforementioned requirement.

To proceed, introduce a second particular value of $\gamma=\gamma'_c \equiv 1.0$.  This value of $\gamma$ by Equation (\ref{gamma}) determines a value $\beta=\beta'_c$, that defines a second critical mass $M'_c$ by Equation (\ref{M0}).  The nature of the equilibrium, or absence of equilibrium, depending on the total mass $M$, can now be distinguished into the ranges : $M>M_c$, $M=M_c$, $M'_c<M<M_c$, and, $M<M'_c$.  

If $M>M_c$ and $\gamma<2/3$, Equation (\ref{H_2}) gives a negative $H_2$ in violation of $H_2$ falling in the range $0<H < \sqrt{\beta}$, so equilibrium does not exist even when we allow for a central mass sheet.  There is too much mass and the heating cannot keep up with the radiational loss.  So the entire mass must collapse to a single mass sheet located at $Y=0$.  

If $M=M_c$, we have $\gamma=\gamma_c=2/3$, $H_2=0$, and the equilibrium solution is everywhere analytical. 

If $M'_c<M<M_c$, $\gamma_c<\gamma<\gamma'_c$, $0<H_2<\sqrt{\beta}$ and the equilibrium solution exists, discontinuous at $Y=0$ where a $m$ mass-sheet is located.  

If $M<M'_c$ and $\gamma>\gamma'_c$, then Equation (\ref{H_2}) gives a $H_2>\sqrt{\beta}$, in violation of $H_2$ falling in the range $0<H < \sqrt{\beta}$.  No equilibrium solution exists, even if mass sheets are allowed, the case of the total mass $M$ being too small for radiational loss to balance against heating.  So the KS slab can only remain in a time-dependent state with temperature increasing with time everywhere.

\subsection{The $M'_c < M < M_c$ Weak Solutions}

To construct an explicit singular solution in the range $M'_c < M < M_c$, apply boundary condition (\ref{theta_bc2}) at $H = \sqrt{\beta}$ ($Y \rightarrow \infty$) to thermal flux ${\mathcal F}$ given by Equation (\ref{thermal_flux_F}): 
\begin{equation}
\label{thermal-flux_layer}
{\mathcal F} (H) = {2 \over 3 \beta^2} \left(\sqrt{\beta} - H \right) \left[ H \left( H + \sqrt{\beta} \right) - \beta \left( 2 - 3 \gamma \right) \right] ,
\end{equation}
\noindent
requiring $H_1 = -\frac{1}{3} \beta^{3/2}( 2 - 3 \gamma)$.  This thermal flux vanishes at $Y \rightarrow \infty$ but is dependent on $\gamma$ and generally not zero at $Y = 0$, the point corresponding to $H = H_2$.  Figure 7 displays the functional forms of ${\mathcal F}(H)$ constructed, for several representative $\gamma$ values.  We call these functional forms unmodified, having analytically extended each of them from the physical 
range of interest $0 < H <   \sqrt{\beta}$ into the negative $H$-range.

The $\gamma = \gamma_c = 2/3$ thermal flux is unique, with $H_2=m=0$ and satisfying boundary conditions (\ref{theta_bc1}) and (\ref{theta_bc2}) at $H = 0,  \sqrt{\beta}$.  So ${\mathcal F}(H)$ is physical and defines an infinity of continuous solutions in the full range $-\sqrt{\beta} < H <   \sqrt{\beta}$, ie, $-\infty < Y <  \infty$.  These solutions all have $M=M_c$ differing in the central temperature $\theta(0)=\theta_1$.  

For $\gamma_c < \gamma  <\gamma'_c$, the solution is singular, obtained by first truncating off the function ${\mathcal F}(H)$ at the limit $H=H_2$ of the range $H_2 < H < \sqrt{\beta}$ that is now to be identified with $0 < Y < \infty$, for $H_2$ given by Equation (\ref{H_2}).   In Figure 7, the lines $H = H_2$ are drawn from the abscissa to the respective ${\mathcal F}(H)$ curves in the range $\gamma_c < \gamma  <\gamma'_c$.
The modified ${\mathcal F}(H)$ located in the interval $-\sqrt{\beta} < H < -H_2$, identified with $-\infty < Y < 0$, is defined by the transformation ${\mathcal F}(-H) = -{\mathcal F}(H)$ as illustrated in Figure 8.  These two continuous solutions in $Y > 0$ and $Y < 0$ are assured to 
match the cold, discrete sheet of mass $m$ related by Equation (\ref{m}) to the jump of $2 H_2$ in $H(Y)$ across the sheet.  

A weak solution for the $\gamma=0.7$ is shown in Figures 9.   This solution is obtained by integrating Equation (\ref{thermal_flux_F}) with respect to $H$ for $\theta(H)$, using Equation (\ref{thermal_flux_H}), over the range $H_2<H<\sqrt{\beta}$.  From it we obtain the symmetric temperature $\theta(Y)$ in $0<|Y|$ up to a free central temperature that we set to zero, ie, $\theta(0)=\theta_1=0$, where the cold sheet of mass $m$ is located.  However, this zero temperature is associated with a finite thermal conductive flux entering the sheet from both sides, in evidence in Figure 8 showing ${\mathcal F} \ne 0$ at $H=H_2$.  In other words, $\theta(Y)$ goes to zero on either sides with an infinite gradient such that ${\mathcal F}$ is finite and nonzero at $Y=0$.  

The central sheet should not be thought of being cold with zero temperature but as the limiting case of a layer cooling to zero temperature, and collapsing during an evolution to achieve equilibrium.  The collapse proceeds with density $D$ becoming a Dirac delta function to account for the mass $m$ and with $P$ holding steady at unity.  The right hand side of energy-balance equation (\ref{energy2}) integrated over the thickness of the sheet gives a constant in time, the steady net loss of energy by the collapsing sheet balanced by the steady inward thermal fluxes on both sides of the sheet.  During such a collapse, $D$ increases in the sheet to become a Dirac delta function, temperature decreases to keep $P = 1$, producing a steady net loss.  This is the nature of the $2/3 < \gamma < 1$ equilibrium states, in each case taking the macroscopic physics of Equation (\ref{energy2}) to remain valid for all time in the interior of the mass sheet.   In the next Section an example is given for a collapsing sheet that has a distinctly different energy process from the exterior continuous parts of the SK-slab.

\section{The interior physics of a discrete mass sheet}

To complete our study, we present two alternative interior-models for a mass sheet matching an exterior continuous solution of our KS slab.  As a sheet collapses with decreasing temperature and unbounded growth in density at a fixed total mass $m$, we expect the optically-thin assumption to break down.  The nature of heating would also be changed.  Heating in the prominence is poorly known or understood, which is the reason for adopting the ad hoc form given by Equation (\ref{heating}), used in earlier prominence modeling works.  

For illustration purpose we consider the case of a finite-thickness mass sheet whose interior is optically thick to all radiation, subject to no heating, but maintained by thermal conduction to be isothermal, allowing for the uniform temperature to be as low as it happens to be at the interface with the exterior region.  In the exterior, described by our KS-slab solutions, the total heat generated is exactly balanced by the total radiative loss. No thermal flux enters mass sheet.  When this sheet is at a sufficiently low temperature, the degree of ionization would be so low that the frozen-in condition must break down.  Then, resistive flows of weakly ionized fluid across the "supporting" magnetic field can take place, the second alternative interior-model that takes us a step closer to the observed prominence with its descending threads and rising bubbles.    

\subsection{Composite KS-slabs}

Consider the case $\gamma = 9/16$ in Figure 7, as an example of $M_c < M$, i.e., $\gamma < 2/3$.
The unmodified thermal flux ${\mathcal F}(H)$ has two zeros in the range $0 < H < \sqrt{\beta}$, one at $H = \sqrt{\beta}$, imposed by the boundary condition at $Y \rightarrow \infty$ and the other at a point at $0 < H = H_3 < \sqrt{\beta}$.   This means that the heat generated in the outer layer $H_3 < H < \sqrt{\beta}$ is entirely radiated away in that layer.  In other words, there is enough mass in the system to create an outer layer that is by itself in global energy balance.  This layer provides no energy to be transferred by thermal conduction to the inner layer, whatever the latter may be.  Integrating Equation (\ref{thermal-flux_layer}) we obtain $\theta(H)$ up to a free constant $\theta(H_3)=\theta_1$, the temperature at the interfaces $H=\pm H_3$.   In a time-dependent cooling of a $M_c < M$ slab, depending on the nature of its initial state, it is possible for such an outer layer to first form and and then stabilize, symmetrically placed on the two sides of $Y = 0$, filling up the regions $|H| > H_3$.  The rest of the mass $m$ in $|H| < H_3$ may then exist in two forms which we now specify.

If we define the mass $m$ as a collapsed, discrete mass-sheet at zero temperature, we have $m g = {B_0^2 \over 2 \pi} H_3$ in analogy to Equation (\ref{m}).  Then, the monotonically increasing $B_z=B_0H(Y)$ jumps from $-H_3$ to $H_3$ across $Y=0$, with the point $Y=0$ mapped into the interval $-H_3<H<H_3$ and the intervals $0<Y^2$ into the iintervals $H_3^2<H^2$.  We must set the boundary condition $\theta(\pm H_3)=\theta_1=0$ for the external solution.  A static solution of this kind with $\gamma = 9/16$ is displayed in Figure 10.  The energy equation  (\ref{energy2}) applies to the exterior solution but not to the interior of the sheet.  

The mass $m$ can also exist as a symmetric, isothermal slab of a prescribed temperature $\theta_1$, of a finite width occupying $-Y_3<Y<Y_3$, matched continuously into a pair of external solutions of our KS-slab occupying symmetrically in $-\infty<Y<-Y_3$ and $Y_3<Y<\infty$ subject to the boundary conditions that the exterior temperature $\theta=\theta_1$ at $Y=\pm Y_3$.  First prescribe $\theta_1$ and pick up the value of $H_3$ from Figure 8 for $\gamma = 9/16$.  The exterior temperature $\theta(H)$ in $H_3^2<H^2<\beta$ is then determined.  The picked value of $H_3$ determines $m$, sufficient for constructing an isothermal solution that locates $H=\pm H_3$ at a pair of points $Y=\pm Y_3$ in physical space.  This isothermal solution then self-consistently matches the outer solutions at $Y=\pm Y_3$ with $\theta(\pm H_3) =\theta_1$ and ${\mathcal F}(\pm H_3)=0$.  The construction is then complete, giving continuous distributions $P(Y)$, $D(Y)$, $\theta(Y)$ and $H(Y)$ over $-\infty<Y<\infty$.  Figures 11, 12, 13, 14 display 4 solutions computed in this way, all sharing the same $M$ and $m$ but for a set of progressively smaller temperatures $\theta_1$ as indicated.   Although a dynamical time-dependent theory is outside the scope of our study, our examples mimic the cooling of the central core of a fixed mass $m$, holding the total mass $M$ also fixed.  As the central isothermal core cools quasi-steadily, that core collapses into a slab of diminishing width, approaching a Dirac delta-function in the limit of $\theta_1 \rightarrow 0$.  Such a collapse proceeds first out of its optically-thin state into an optically-thick state, and then to extremely low temperatures when the assumptions of high ionization and the frozen-in condition must fail.
    
\subsection{Resistive vertical cross-field flows}

In an extremely cold and dense fluid, the electrical conductivity cannot be taken as infinite by two effects, the low degree of ionization and the narrowing of collapsing layer to such a small size that the weak but finite electrical resistivity has become important.  If resistivity is significant, the Kippenhahan-Schl\"{u}ter slab can be extended to include a steady vertical resistive flow.
Consider the single-fluid resistive hydromagnetic equations (\ref{momentum}) and (\ref{induction_resistive}),
taking the resistivity $\eta$ uniform in space for simplicity.  In the presence of the simple steady vertical velocity
\begin{equation}
\label{velocity}
{\bf v} = \left[0, 0, v(y)\right],
\end{equation}
\noindent
the rate of change of momentum on the left hand side of Equation (\ref{momentum}) vanishes identically and 
we recover the force-balance equation (\ref{fb}).  We also recover the steady energy transport equation (\ref{energy}).  The mass conservation is automatically satisfied.  Therefore, each of the magnetostatic solutions investigated in this paper is consistent with the above steady, vertical, resistive flow with $\eta \ne 0$.

The induction equation (\ref{induction_resistive}) reduces
to a simple second-order ODE which can be integrated once to give
\begin{equation}
\label{velocity_2}
v(y) = -\frac{\eta}{L_0} {dH \over dY} + v_0 ,
\end{equation}
\noindent     
where $v_0$ is an integration constant and $L_0$ is the density scaleheight.  In all our solutions $\frac{dH}{dY} \rightarrow 0$ as $Y \rightarrow \infty$. 
Therefore, we set $v_0 = 0$ for a system static in the far regions.  Combined with magnetostatic equation (\ref{density}), we have the velocity
\begin{equation}
\label{velocity_3}
v(y) = -\frac{\eta \beta}{2 L_0} D .
\end{equation}
\noindent      
In this resistive flow the energy released from current dissipation is conserved by the release of gravitational potential energy associated with the steady fall of matter.  The prominence-drainage event of Liu, Berger \& Low (2012) is thus interesting.  The drainage liberates gravitational potential energy.  If that drainage is across field lines, the liberated energy manifests, not as an increase in kinetic energy, but as dissipative energy that fuels the radiative loss of the prominence plasma.  

Suppose $\eta$ is small but not zero.  Then to a first approximation we neglect $\eta$ and $v$.  The results of Section 3 have the following significance.  Suppose $M>M_c$, static equilibrium is achieved in a relaxation by withdrawing a specific portion $m$ of the given total mass $M$ into a collapsed central sheet.  This process implies that the density $D$ grows unboundedly at $Y=0$.  At some point, for any given $\eta$ small but not zero, the velocity $v$ becomes significant at $Y=0$ leading to a breakdown of the frozen-in condition in spite of the assumed smallness of $\eta$.  

For example, consider this resistive flow taking the form  
\begin{equation}
\label{velocity_isothermal}
v(y) = -\frac{\eta \beta}{2 h \theta_1} {\rm sech}^2  {\sqrt{\beta} \over 2 \theta_1} Y  ,
\end{equation}
\noindent 
with $D$ given by Equation (\ref{isothermal_density}) at the uniform temperature $T = \theta_1 T_0$.   We have a vertical down flow that is maximum at $Y = 0$ but  decreasing with $|Y|$ exponentially.  
Use the normalization $T_0 = 10^4 K$ so that $L_0 \approx 300~km$.  The Spitzer electrical conductivity $\sigma = 2 \times 10^7 T^{3/2} ~s^{-1}$ gives a resistivity $\eta \approx 3 \times 10^6 \theta_1^{-3/2}$.  The peak resistive flow at $Y = 0$ is then
of the order of $\frac{\eta \beta}{2 L_0 \theta_1} \approx 5 \times 10^{-2} \frac{\beta}{\theta_1^{5/2}}$.  To produce a flow of the order of $1 ~ km s^{-1}$, would require $\beta$ large and $\theta_1$ small, e.g., $\beta \approx 10^3$, $\theta_1 \approx 10^{-2}$, implying a highly sheared field and low temperature of the order of $T \propto 10^2 K$.  Such an extreme condition is the reason for a long-standing pessimistic perception that {\it steady} resistive slippage of ionized plasmas as the result of the Spitzer conductivity is unimportant.  The spontaneous formation of a mass-sheet singularity is a novel example of how precisely this kind of extreme circumstances may develop in the corona under which resistive slippage does become important.

This effect is more interesting as a recurrent time-dependent process.   If we neglect the weak resistivity altogether, the frozen-in condition demands for an equilibrium state with a singularity.  This singularity generates a resistive flow of an unbounded magnitude until the singularity formation is stopped by current dissipation as a heat source for the cold fluid.  The heating can increase the degree of ionization, restoring the frozen-in condition, whereupon a collapse into a mass sheet is once again initiated.  This time-dependent process is worthy of further investigation.  

\subsection{Ambipolar diffusion in the KS slab}

To address the degree of ionization, a two-fluid description is needed \citep{gilbert02,gilbert07}.  Such a description is necessarily 2D, quite outside the scope of our study, but the following is an interesting closing point to make.  

In a 1D slab of weakly ionized plasma, the dominant neutrals are kept from slipping freely across the field by collisions with the ions that tend to be tied to the field.  The system cannot remain 1D because there will be diffusion of the neutrals in both $Y$ and $Z$ directions.  If we concentrate on the vertical force balance in the central region of a KS-like slab, the following order of magnitude consideration adapted from Spitzer (1978, page 294, eqn. (13-55)) is instructive.  Take the ions to be rigidly tied to the locally horizontal field as the neutrals fall steadily vertically across the field lines in the presence of a mutual frictional force.  For the neutrals as a fluid, its weight is balanced by the upward frictional force acting on it, denoted by $F_{z, i,n}$.  The ions as a fluid experiences the opposite frictional force $F_{z, n,i}=-F_{z, i,n}$ as well as the upward Lorentz force that effectively supports the total weight.  Use the Spitzer approximation,
\begin{equation}
\label{friction_z}
F_{z, i,n} = - F_{z, n,i} = \rho_i \rho_n \frac{c_0}{m_0} v(y) ,
\end{equation}
\noindent  
where $v>0$ is the speed of the neutral fluid falling relative to a stationary ion fluid with densities $\rho_n$ and $\rho_i$, respectively.   The parameter $c_0$ is a thermodynamic constant not greatly sensitive to temperature, ranging in value from $10^{-10}$ to $10^{-9}$ for a hydrogen gas at temperatures in the range $10-10^3K$, and, we recall, $m_0$ is our notation for the mean molecular weight of the fluid which we shall take to be the hydrogen mass.   We also have the vertical balance of forces for the two fluid components:
\begin{eqnarray}
\label{friction_neutral}
\rho_n g = F_{z, i,n} ,  \\
\label{friction_ions}
\rho_i g = F_{z, n, i} + B_0^2 {dH \over dy} .  
\end{eqnarray}
\noindent
Summing these two equations gives the one-fluid Equation (\ref{density}) following the canceling of equal and opposite frictional forces.   

Substituting for the frictional force in Equation (\ref{friction_neutral}) given by Equation (\ref{friction_z}), we obtain, after a canceling of $\rho_n$ as a common multiplicative factor on the two sides of the equation, the speed of fall is independent of $\rho_n$:
\begin{equation}
\label{friction_slip}
v(y) = - {mg \over N_i c_0} ,
\end{equation}
\noindent  
entirely determined by the number density $N_i$ of the ionized fluid.   The speed of fall depends on gravity but it is steady and not of the magnitude of a free-fall speed.   A slippage velocity of $2 ~ km~s^{-1}$ is obtained if the ion number density drops to below $10^5~cm^{-3}$.  Again, this speed of fall would not be significant if not for the principal result of the paper.  If no such poorly ionized fluid form, the frozen-in condition would obtain.  Then, if the mass frozen in the the field is greater than $M'_c$, the collapse of the core of the KS slab would bring the temperature at the center to as low as it takes for the number density of the ions to be reduced to the level where the frozen-in condition is invalid.   Heating is produced by the friction force and the flow can be thus quenched when the ionization level is increased.  What is interesting is that if the frozen-in condition returns, and the total mass still exceeds $M'_c$, cooling and collapse of the optically thick core recur.  The condensation is thus recurrent and not static for these theoretical reasons.

\section{Discussion}

The vertical descent of prominence threads at less than free-fall speeds has been interpreted in several ways.  The large-scale twisted field in the prominence cavity may be evolving with flux-tubes falling bodily together with their frozen-in plasmas as the result of intermittent magnetic reconnections occurring somewhere else along a flux-tube \citep{low05,petrie05,chae10}.  van Ballegooijen and Cranmer (2010) proposed that the prominence plasma is flowing under the frozen-in condition in a 3D tangled global field that has a strong vertically-oriented component.  In contrast, here we propose a spontaneous breakdown of the frozen-in condition to allow a resistive flow across a local horizontally-oriented field.  What seems clear is that all these dynamical processes may occur simultaneously.  

Our KS-slab model shows that the nonlinear coupling of force-balance with energy-balance under a rigorous imposition of the frozen-in condition must generally produce a cold, zero-thickness sheet of mass if static equilibrium is to be attained.  This mass sheet contains a discrete electric current that generates the Lorentz force supporting the weight of the sheet.  In a real plasma of a high but finite electrical conductivity, this discrete current must dissipate resistively.  In other words, the typically high degree of the frozen-in condition must physically break down within the mass sheet.  In a time-dependent relaxation to such an equilibrium, a mass sheet would first develop during the time when the frozen-in condition applies approximately. Then, in that developing mass sheet, the frozen-in condition soon breaks down to produce a downward resistive flow across the supporting field.    Yet, the frozen-in condition is not broken down permanently.  The implied resistive heating in the sheet, fueled by gravitational potential energy released by the falling, poorly-ionized plasma, can bring back the frozen-in condition, but only to recreate the circumstance under which a thin, cold sheet is again inevitable.   This recurrent spontaneous breakdowns of the frozen-in condition is a physically attractive hypothesis to explain the constantly restless interior of a quiescent prominence as revealed by {\it SDO/AIA} and {\it Hinode} observations. 

Gravity, anisotropic Lorentz force, optically-thin radiation, heating, and anisotropic thermal conduction all play essential roles in creating what seems quite general circumstances under which thin collapsed mass sheets must form.  The particular optically-thin radiative loss and heating described by Equations (\ref{radiation}) and (\ref{heating}) are artificial, but they capture basic properties sufficiently well for us to use the KS-slab to demonstrate the inevitability of mass-sheet singularity with mathematical rigor.  Our demonstration is surprisingly complete.  The mass-sheet singularity in our KS-slab is resolvable into a sheet of finite width using an exact solution for a cross-field resistive flow to account for the breakdown of the frozen-in condition.   Both the static and time-dependent problems of this complex coupling are formidable, especially for 2D and 3D systems.  Direct numerical simulation holds the only hope for making progress but to be fruitful this approach must start with a physics-based hypothesis like the one presented here.  Much further work is needed to treat physically more realistic energy processes for applications directly to the prominence \citep{gilbert02,gilbert07} and other astrophysical systems, notably the interstellar medium \citep{maclow95,parrish05}. 

Observation remains the spearhead in the investigation of quiescent prominences \citep{berger08,berger10,berger11,chae10,gaiz98,heinzel08,labrosse10,lites10,mackay10,martin94,martin10,th95,lites10,okamoto07,okamoto08,okamoto10,labrosse10,mackay10,vanB04}.  Presently not all the observed properties in prominences can be theoretically understood or understood in a unified way.  For example, the sub-arcsecond threads, revealed by ground-based, high-resolution, static $H_{\alpha}$ images of on-disk prominences, have remained difficult for theory to unify with the space observations of prominences at the solar limb from {\it Hinode} and {\it SDO/AIA} \citep{zirker98,lin03,lin05,su11}.   These two views of prominence structures have also not been reconciled with the polarimetric observations of prominence and coronal magnetic fields \citep{leroy89,casini03,casini99,judge06,lopez03,tomczyk08,trujillo02,dove11}.   

Observations can pose good physical questions to be addressed theoretically in their own rights, without necessarily aiming at a comprehensive explanation of all the rich observations.  In the face of many unanswered questions, our series of paper seeks to build up our intuition for seeing the complex observations in terms of the elementary physical principles.   Previous works of this kind have addressed magnetic support of the prominence weight under the frozen-in condition using 2D and 3D field models with energy transport greatly simplified or ignored for the sake of tractability \citep{aulanier02,chae10,low95,low04,low05,lopez06,manchest04,petrie05,priest89,regnier04,vanB04,wu87}.  In complement, models of 1D sophisticated energy transport along field lines, prescribed to be fixed in space, made progress with insights into the condensation processes by suppressing the dynamical interaction of the field with energy transport\citep{karpen06,karpen08,luna12}.   Heasley and Mihalas (1976) pioneered studying the coupling of radiative transfer to the equilibrium forces in a slab model.   Allowing for gravity and a heuristic approximation of the Lorentz force, this interesting work encountered an inevitable presence of collapsed mass-sheets, perhaps a rudimentary form of the singularity we found in our KS-slab model; see also Anzer \& Heinzel (1999).  

Collapsed mass sheet may be a common occurrence in the corona.   All it takes is a failure to bring energy to where it is needed whenever a particular mode of energy transport becomes over constrained by the coupled requirement of balancing the forces in the gravitationally stratified fluid.  Let us conclude with the following intuitive physical picture that has emerged from our study.

In our KS-slab, there are two thermal-insulation mechanisms that can keep the prominence cool against the million-degree hot, thermally highly-conducting corona.  The magnetic field is an insulator against thermal conduction across it.  Not so well recognized is the self-insulation of the prominence plasma along a flux tube.  In a sheared U-shaped flux tube, the frozen-in mass is gravitationally stratified along the tube so as to progressively radiate away the two thermal fluxes conducted inward from the high-temperature upper parts of the tube in the corona.  If the total mass is large enough, the steady state may be one in which these thermal fluxes are depleted before the bottom of the tube is reached.   Then the bottom contracts into a cold dense mass simply from want of heat.  The degree of ionization of such a core would fall below a threshold whereupon the frozen-in condition must break down.   

The prominence/corona transition is similar to the corona/chromosphere transition of the solar atmosphere but that analogy is just imperfect.  In the solar atmosphere, the bottom of the chromosphere never cools below about $5000^oK$ because the photosphere beneath it has no bottom and is heated actively by the solar luminosity coming through from below.  If the prominence plasma sitting at the bottom of an U-shaped tube has no internal heat source and is self-insulating, it would simply cool to an equivalent of the solar photosphere with a very different property, one with such high density and low temperature as to break the frozen-in condition.  This massive core would then flow resistively across the field into the adjacent flux tubes.  No steady flow is expected, just a recurrent process of the breakdown alternating with restoration of the frozen-in condition by resistive heating.  

In realistic 3D situations, the super-cooled core of a prominence plasma slips resistively out of its optically-thin chromospheric cladding into an adjacent flux tube.  There it can either continue to slip resistively into other tubes further away or else become so ionized in the new hot environment as to be frozen into the new tube it has just occupied.   A sufficiently massive super-cooled core can thus fall through the bottoms of successive U-tubes leaving behind in each of these tubes ionized plasmas at chromospheric temperatures, until the former runs out of mass to remain a collapsed core.  A vertical thread of chromospheric plasma is naturally produced.  The movie of such a process made at the cool chromospheric temperature would then show such a thread forming from out of nowhere with a general downward descent.   The presence of U-shaped flux tubes is the pre-requisite of this process.

Although such massive, poorly ionized cores are time-dependent objects, this picture argues for the presence of prominence plasmas that are more massive, more dense, much colder, and less ionized than currently being contemplated.  Perhaps the canonical observation-based numbers describing the prominence, electron density of $10^{10-12}~cm^{-3}$, temperature of $\approx 5000^{o}K$, degree of ionization $0.1 - 10$\%, are {\it only} the properties of the optically-thin, chromospheric cladding of the not yet observed, transient, optically-thick cores.  The tangled-field models of van Ballegooijen and Cranmer (2010) readily give peak densities well in excess of the aforementioned canonical values.  These models assume a moderate field of only 10G, whereas even larger densities are implied for fields as high as 60G given by polarimetric observations \citep{casini03,lopez03}.  Our KS-slab suggests that even more extreme densities might occur commonly in prominences.   

We emphasize that upward and downward motions accompany this process.  The unloading of a flux tube to give mass to the one below results in the former rising by its resultant overcompensating Lorentz force.  Correspondingly, the mass-receiving tube below sinks heavily until the Lorentz force of its stretched field can hold it swaying into position.  Here is an instance of a flux-tube of fluid rising or falling bodily, carrying its frozen-in magnetic flux along with it.  

Our hypothesis suggests that the cavity magnetic field supporting the prominence is "porous" in spite of the high electrical conductivity.  The observations of Liu, Berger \& Low (2011) thus have the significance as a quantitative statement of how porous that magnetic field can be, namely, that an order of magnitude more mass than retained by the prominence at any one time can pass resistively and agitatedly down the cavity field in a day.  

In the second paper we will get a clearer picture of the role of the magnetic field in this process \citep{low12}.  The field actually is the dominant dynamical player, for it can, on its own, create and restore the frozen-in condition in a permanent restlessness \citep{parker94,janse10a,janse10b}.  This mechanics may be basic to the proposal that the long-lived macroscopic prominence is a form of magneto-thermal convection in the corona proposed by Berger et al. (2011). 

We thank Yuhong Fan, Michael Goodman, Phil Judge and Bruce Lites for discussions.  The National Center Atmospheric Research is sponsored by the National Science Foundation.  TEB was supported by the Solar-B FPP contract NNM07AA01C at Lockheed Martin.  WL is supported by NASA SDO/AIA contract NNG04EA00C.

\begin{figure}
\centerline{\includegraphics[width=120mm]{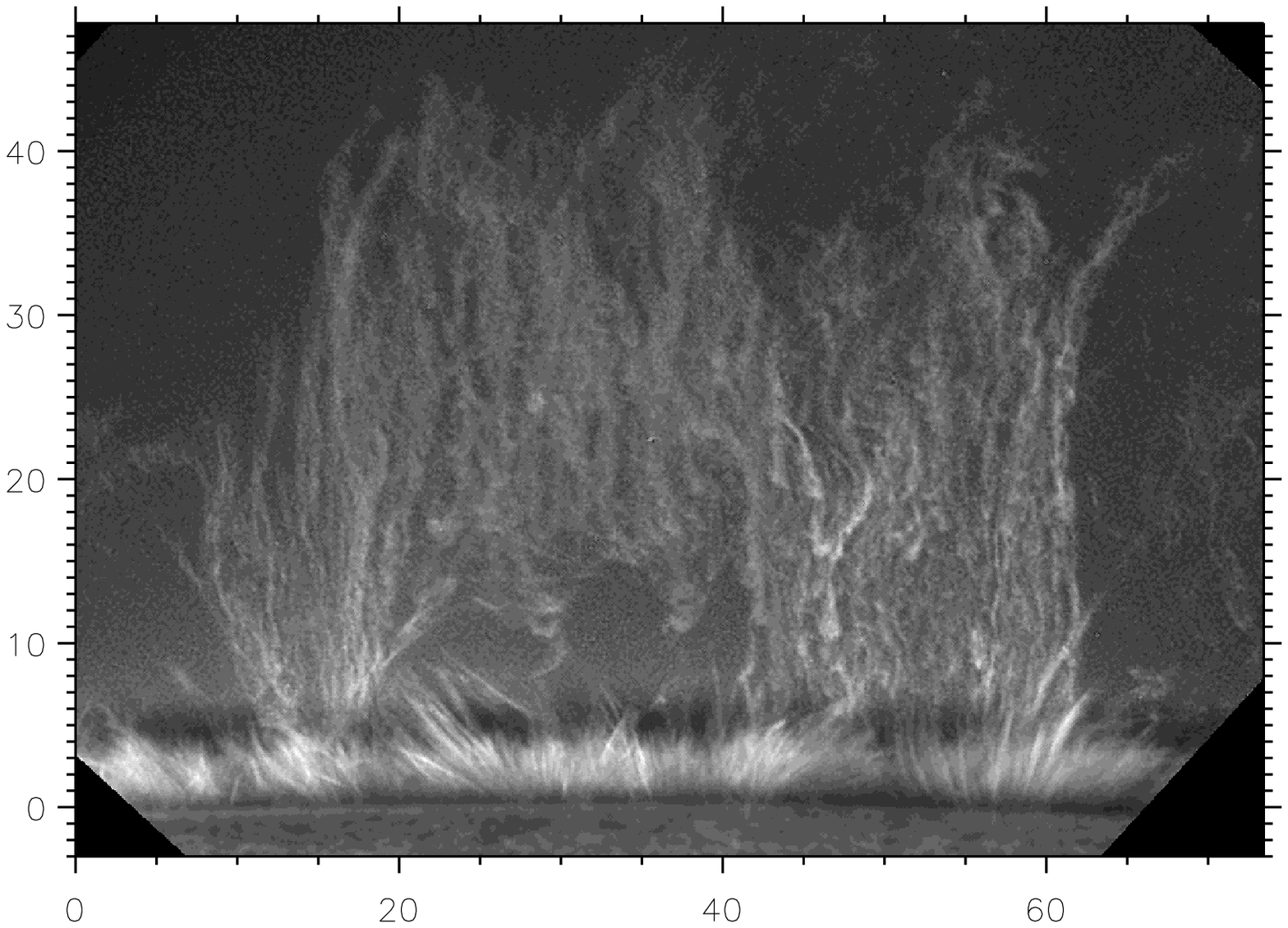}}
%\centerline{\includegraphics[width=130mm]{Fig1.eps}}
\caption{\small{A {\it Hinode}/SOT image in CaII H emission of the fine-scale vertical threads in a prominence observed on October 3, 2007 at the solar limb.}}
\end{figure}

\begin{figure}
\centerline{\includegraphics[width=120mm]{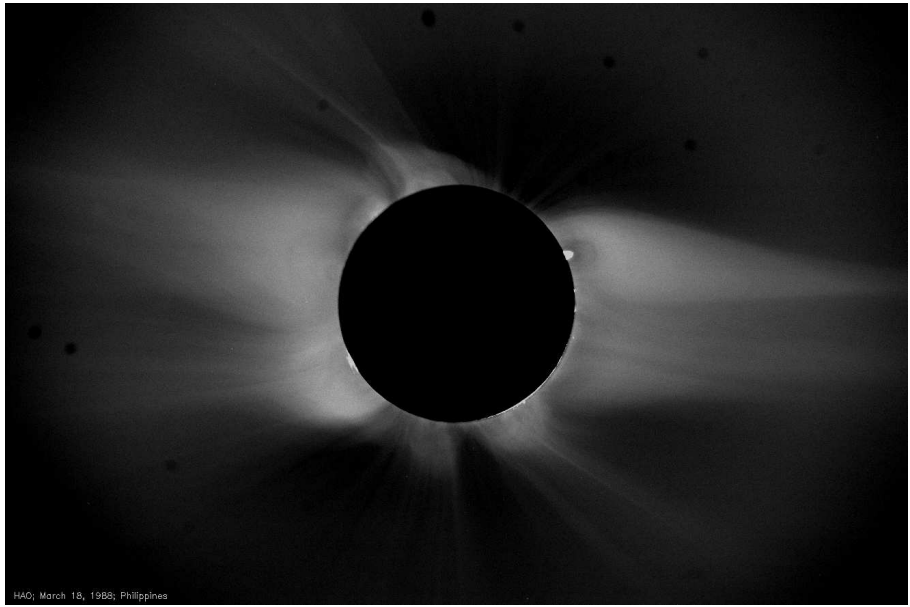}}
%\centerline{\includegraphics[width=130mm]{Fig2.eps}}
\caption{\small{White-light total eclipse of March 18, 1988 showing a well-formed, bright coronal helmet-streamer extending from the north-east limb, with a localized, low-density cavity at the helmet base.  Within that cavity a spatially-unresolved, quiescent prominence appears as a bright blot.}}
\end{figure}

\begin{figure}
\centerline{\includegraphics[width=120mm]{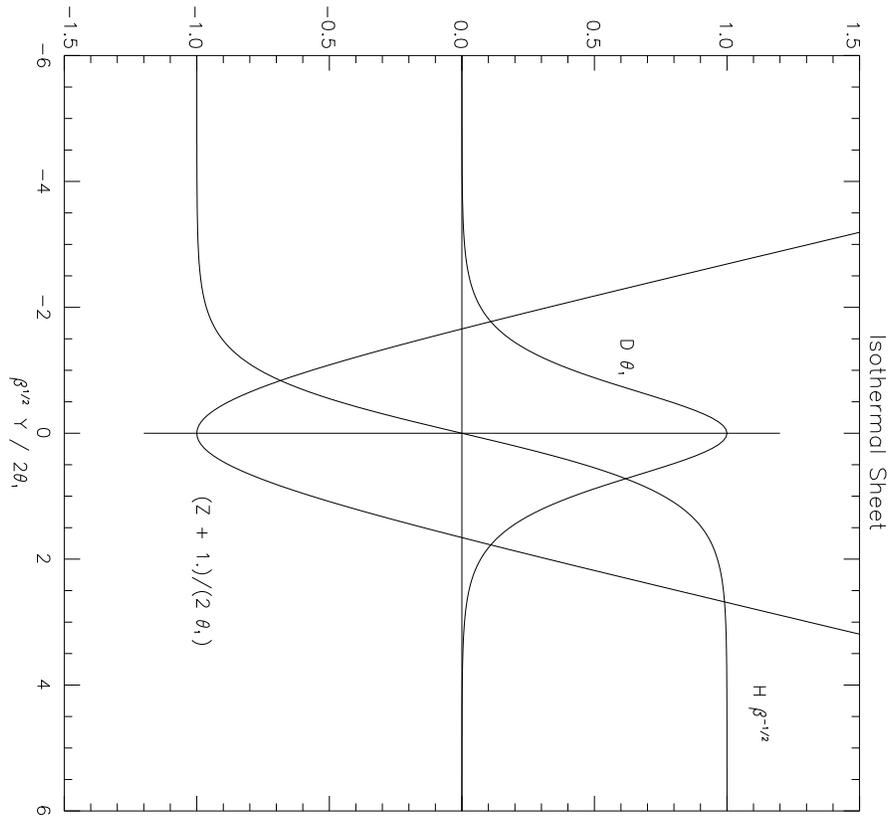}}
%\centerline{\includegraphics[width=130mm]{Fig3.eps}}
\caption{\small{The isothermal Kippenhahn-Schl\"{u}ter prominence slab displayed with graphs of density $D(Y)$; vertical field component $H(Y)$; and, the shape of the bowed field line projected on the $Y-Z$ plane described by $Z(Y)$, as described in the text.
 }}
\end{figure}

\begin{figure}
\centerline{\includegraphics[width=120mm]{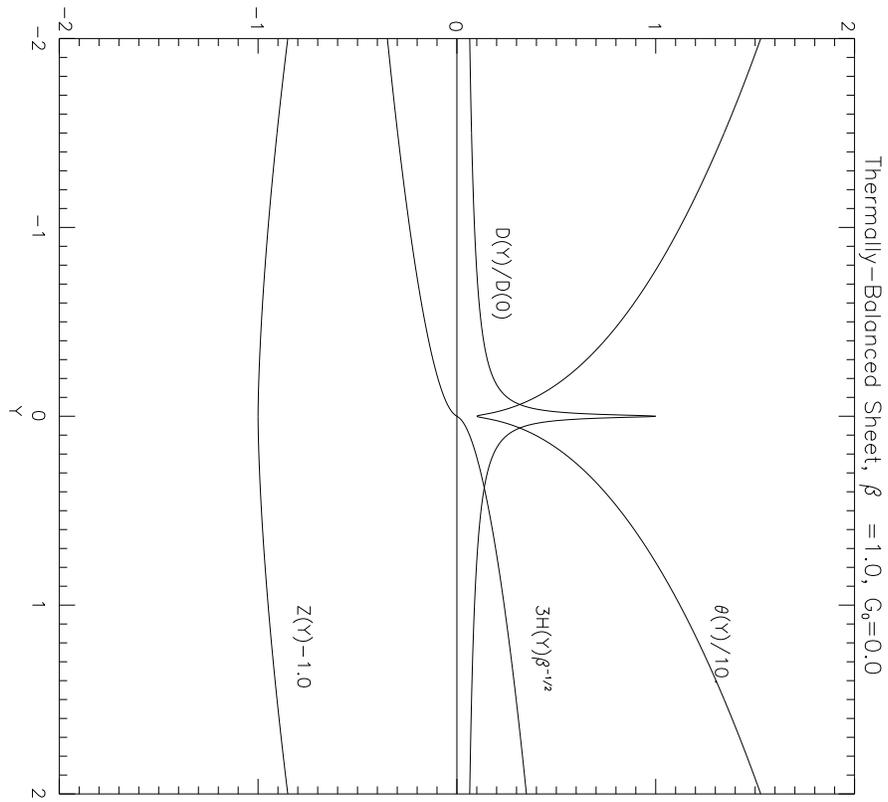}}
%\centerline{\includegraphics[width=130mm]{Fig4.eps}}
\caption{\small{Thermally-balanced continuous solution with $\beta=1.0$, $G_0=0$, showing the sharply spiked density $D(Y)$; rapid rise of temperature $\theta(Y)$ from a low value at the origin; the vertical field component $H(Y)$; and the shallowly bowed field line represented by $Z(Y)$ in the $Y-Z$ plane.  }}
\end{figure}

\begin{figure}
\centerline{\includegraphics[width=120mm]{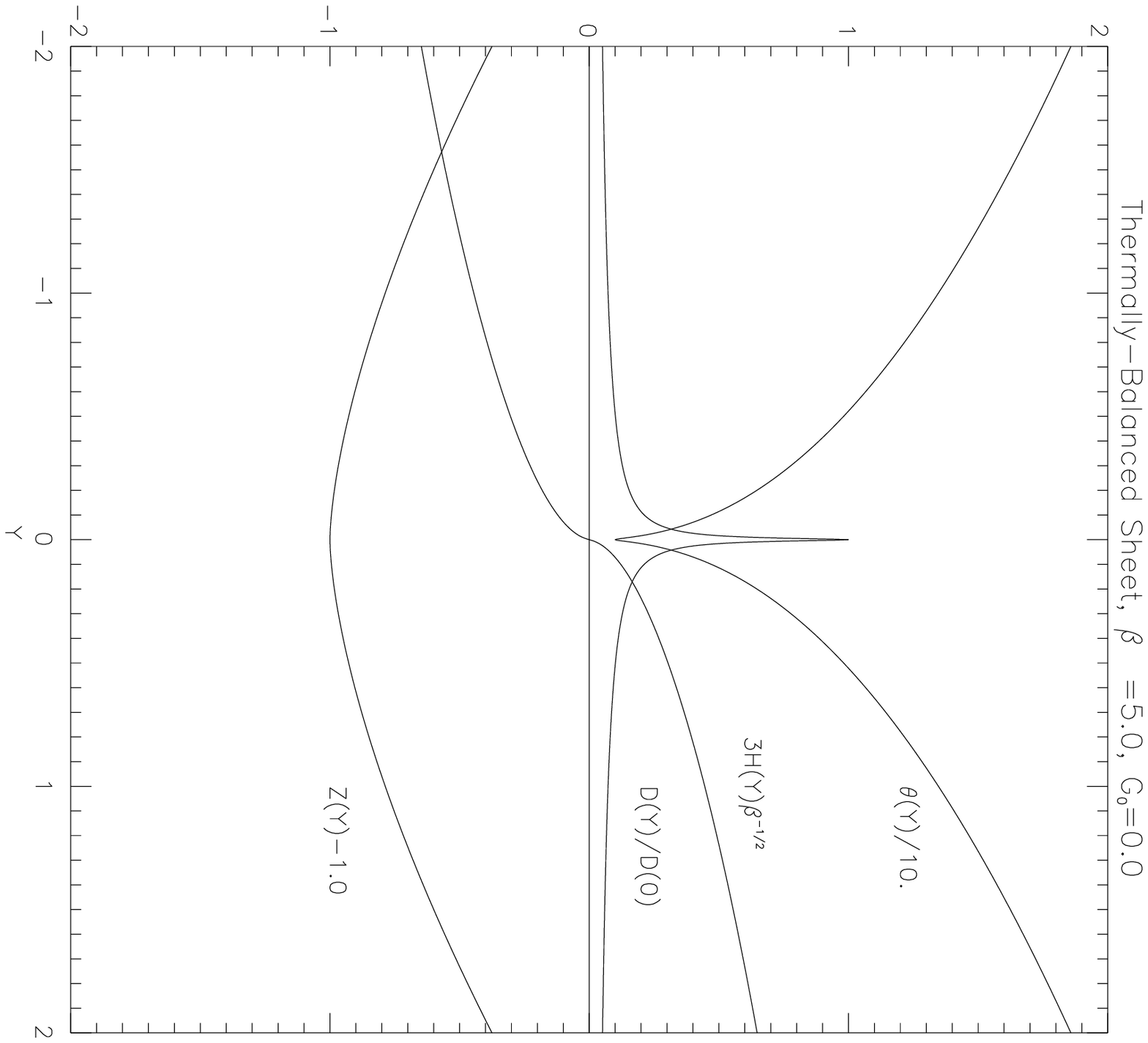}}
%\centerline{\includegraphics[width=130mm]{Fig5.eps}}
\caption{\small{Thermally-balanced continuous solution with $\beta=5.0$, $G_0=0$, in the same format as in Figure 4.}}
\end{figure}

\begin{figure}
\centerline{\includegraphics[width=120mm]{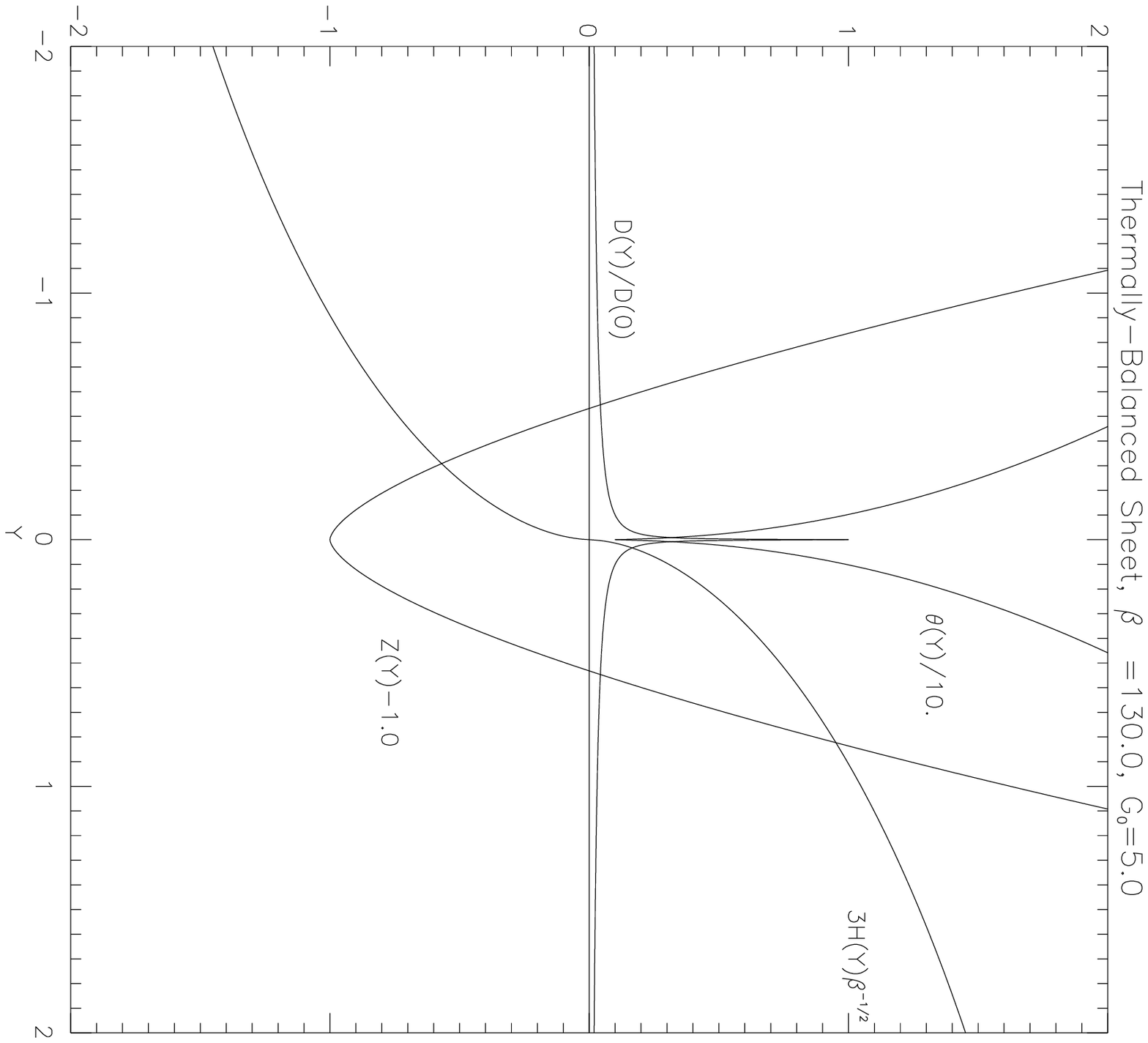}}
%\centerline{\includegraphics[width=130mm]{Fig6.eps}}
\caption{\small{Thermally-balanced continuous solution with $\beta=130.0$, $G_0=5.0$, in the same format as in Figure 4. }}
\end{figure}

\begin{figure}
\centerline{\includegraphics[width=120mm]{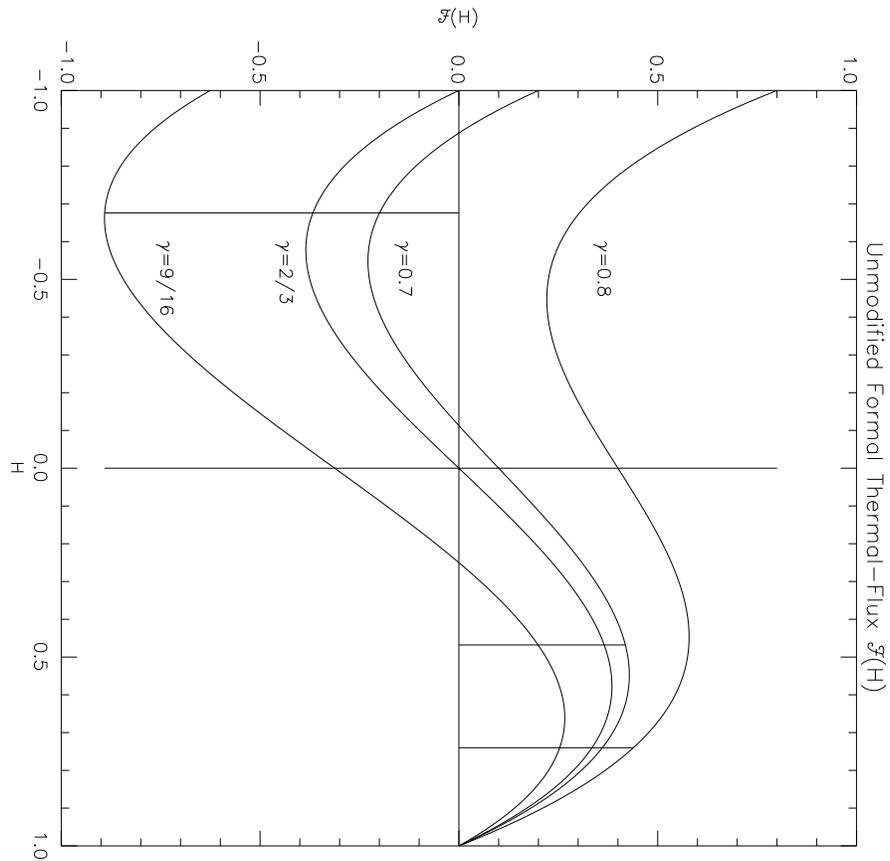}}
%\centerline{\includegraphics[width=130mm]{Fig7.eps}}
\caption{\small{Unmodified thermal flux ${\mathcal F}(H)$ for different values of $\gamma$ indicated over the range $-\sqrt{\beta} < H < \sqrt{\beta}$.  Except for the antisymmetric $\gamma = 2/3$ graph, vertical lines $H = H_2$ are drawn from the abscissa to the respective curves as described in the text. }}
\end{figure}

\begin{figure}
\centerline{\includegraphics[width=120mm]{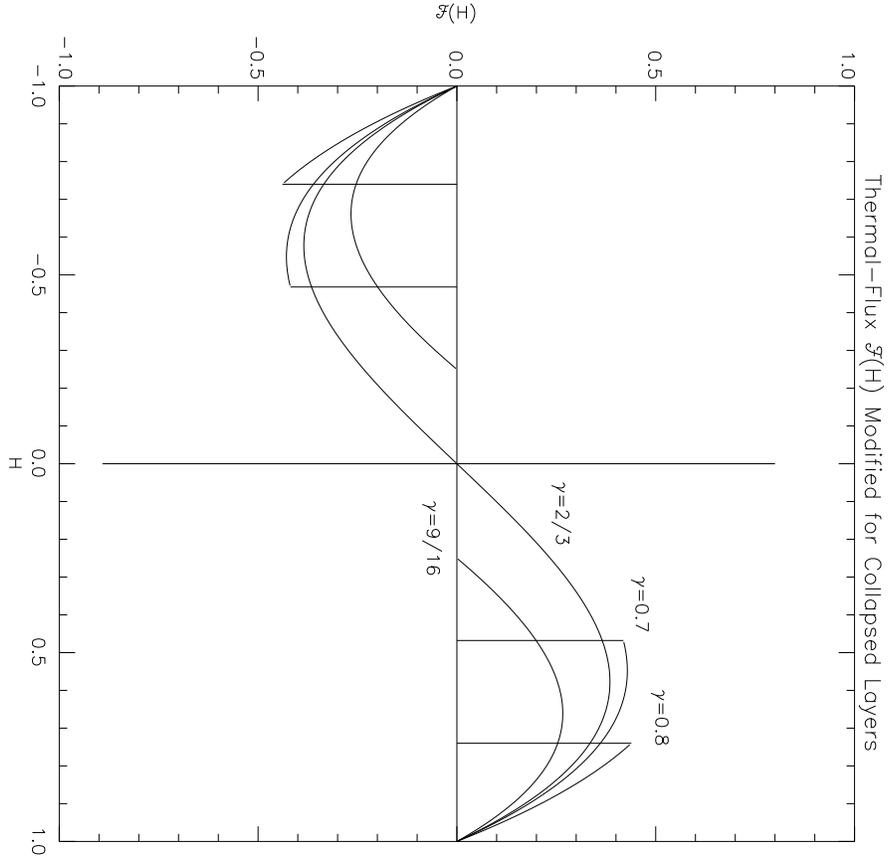}}
%\centerline{\includegraphics[width=130mm]{Fig8.eps}}
\caption{\small{Modified thermal flux ${\mathcal F}(H)$ for different values of $\gamma$ indicated over the range $-\sqrt{\beta} < H < \sqrt{\beta}$.  Except for the antisymmetric $\gamma = 2/3$ graph, the modified ${\mathcal F}(H)$ in $H > 0$ is defined only for $H > H_2$ truncating off the $-\sqrt{\beta} < H < H_2$ portion of the unmodified ${\mathcal F}(H)$ shown in Figure 7.  The modified ${\mathcal F}(H)$ in $H < 0$ is then defined for $-\sqrt{\beta} < H < - H_2$ by an antisymmetric reflection of the part so constructed for $H > H_2$. }}
\end{figure}

\begin{figure}
\centerline{\includegraphics[width=120mm]{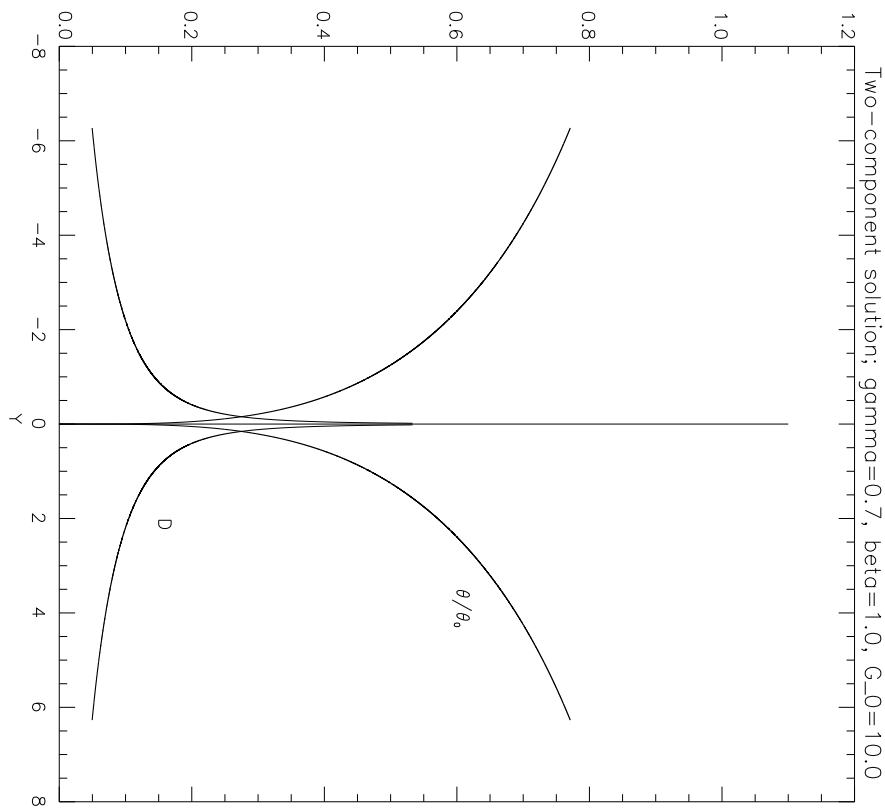}}
%\centerline{\includegraphics[width=130mm]{Fig9.eps}}
\caption{\small{The density $D(Y)$ with a central Dirac delta function and temperature $\theta(Y)$ of a $\gamma = 0.7$ thermally-balanced solution.  The delta function describes a collapsed mass sheet corresponding to the jump of $2H_2$ across the origin of the vertical field component $H(Y)$.  The thermal flux ${\mathcal F}$ for $\gamma = 0.7$ shown in Figure 8 indicates that $\theta(Y) \rightarrow 0$ with an infinite gradient as $Y \rightarrow 0$, delivering a non-zero thermal flux into the mass sheet as described in the text. The constant $\theta_0$ is only a normalization to fit $D(Y)$ and $\theta(Y)$ into the same graph.}}
\end{figure}

\begin{figure}
\centerline{\includegraphics[width=120mm]{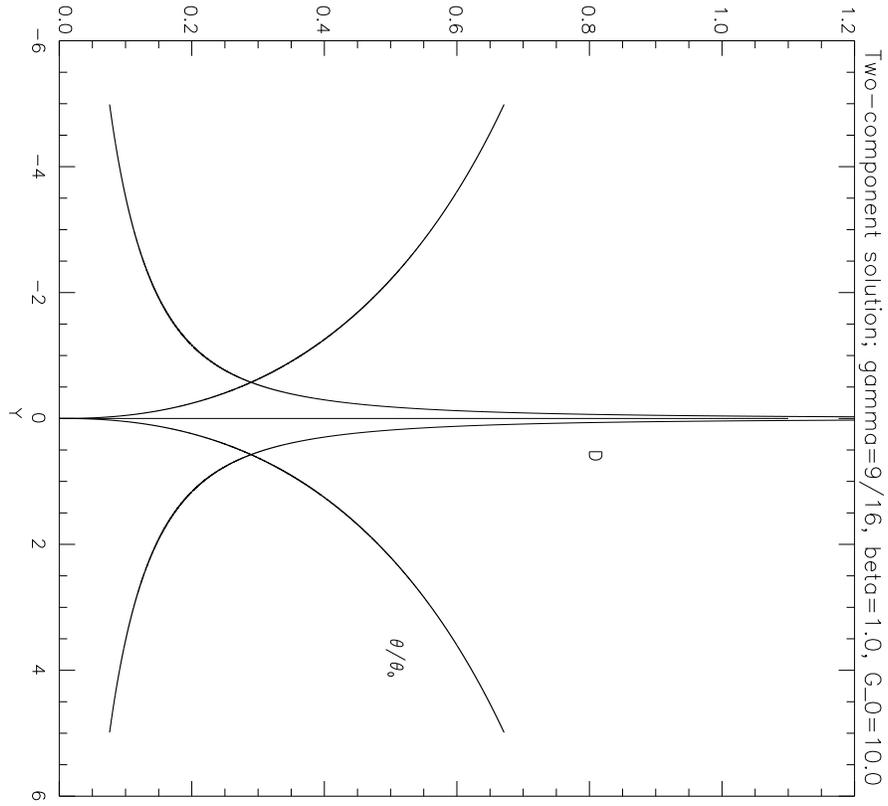}}
%\centerline{\includegraphics[width=130mm]{Fig10.eps}}
\caption{\small{A $\gamma = 9/16$ thermally-balanced solution with a collapsed mass sheet in the same format as in Figure 9 except that, not resolved at the scales of the figure, $\theta \rightarrow 0$ with zero thermal flux at $Y = 0$ as discussed in the text.  In the continuous part of the solution,the total heat input  is balanced exactly by the total radiative loss.}}
\end{figure}

\begin{figure}
\centerline{\includegraphics[width=120mm]{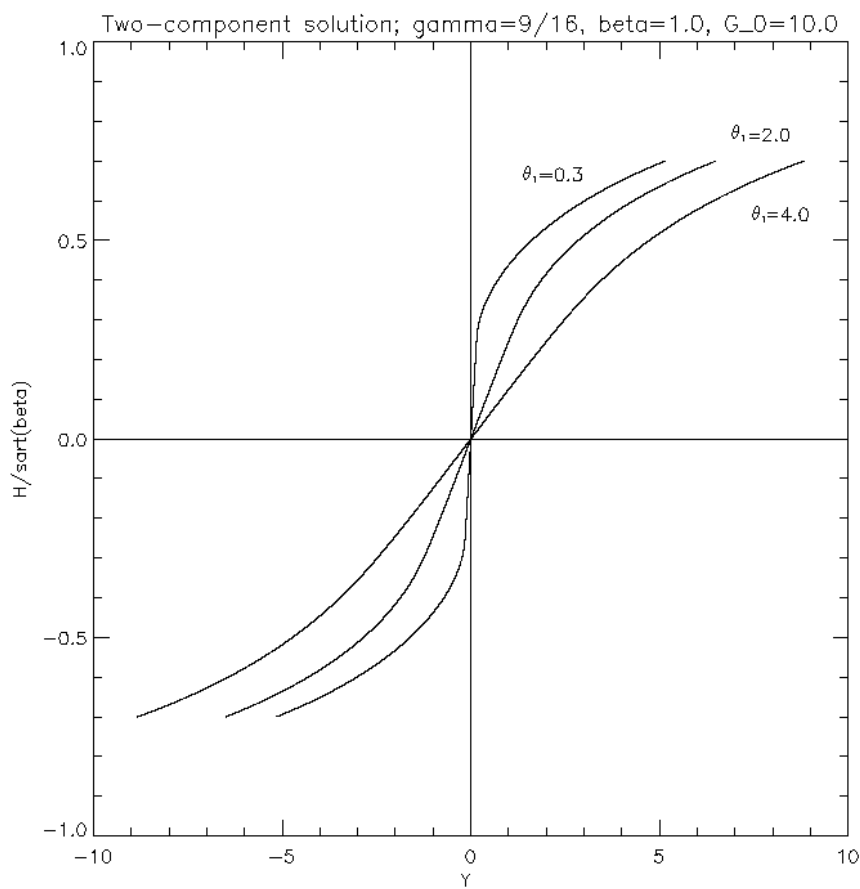}}
%\centerline{\includegraphics[width=130mm]{Fig11.eps}}
\caption{\small{The distributions of the vertical field component $H(Y)$ of the two fluid-component thermally-balanced solutions displayed in Figures 12, 13, 14, distinguished by the central temperatures $\theta_1$ of these solutions as indicated, measured in a common unit.  Each solution is everywhere continuous, having the same total mass $M$ and thus the same range $-\sqrt{\beta}  < H(Y) < \sqrt{\beta}$, with an isothermal solution of temperature $\theta_1$ in the range $-H_3 < H < H_3$ and a $\gamma = 9/16$ thermally-balanced exterior solution in $H_3 < |H|$.  The parameter $H_3$ is the same constant for all three solutions, as described in the text.}}
\end{figure}

\begin{figure}
\centerline{\includegraphics[width=120mm]{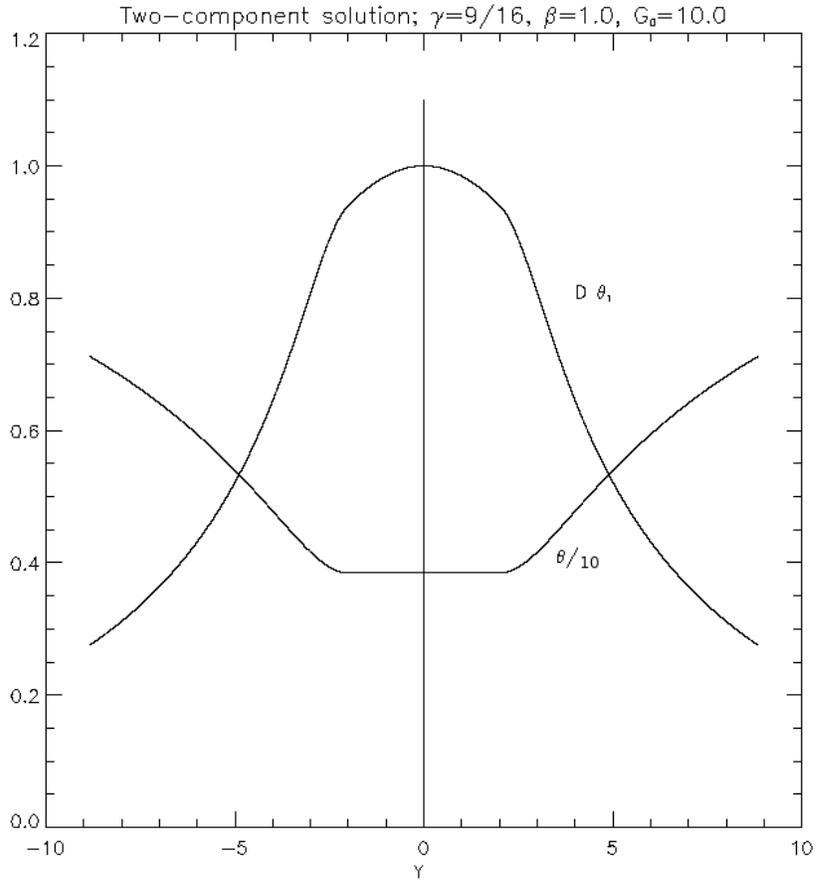}}
%\centerline{\includegraphics[width=130mm]{Fig12.eps}}
\caption{\small{The broad distributions of density $D(Y)$ and temperature $\theta(Y)$ with central temperature $\theta_1 = 4.0$.  The latter shows clearly the location, $-H_3 < H < H_3$, of the inner fluid component by its spatially uniform temperature that continues into a $\gamma = 9/16$ thermally-balanced exterior solution located in $H_3 < |H|$.}}
\end{figure}

\begin{figure}
\centerline{\includegraphics[width=120mm]{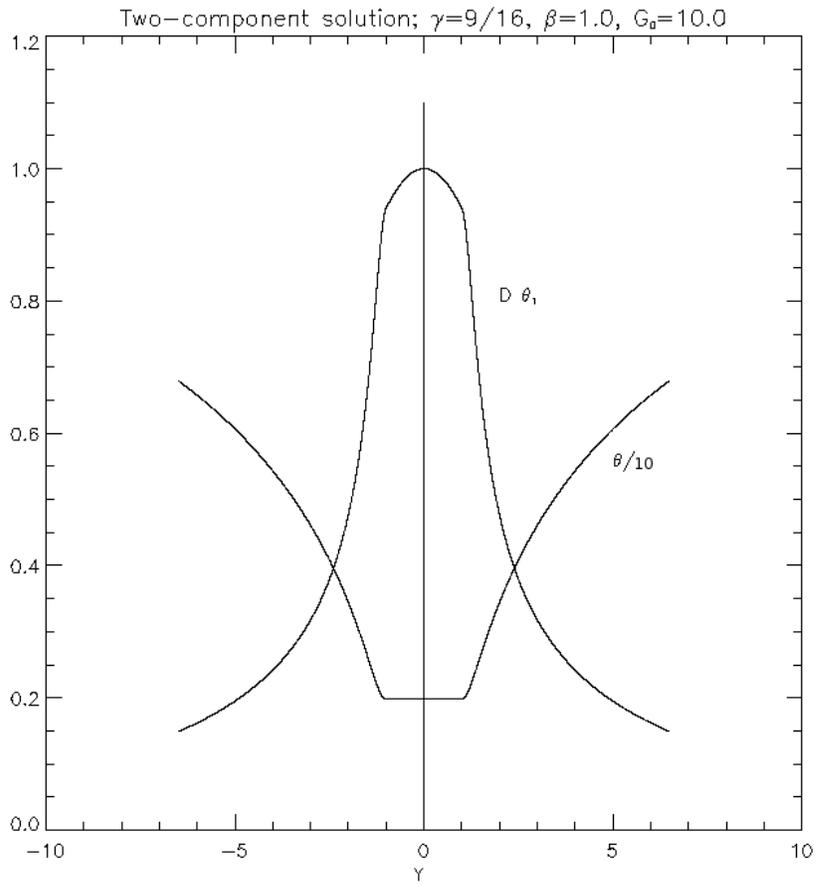}}
%\centerline{\includegraphics[width=130mm]{Fig13.eps}}
\caption{\small{The distributions of density $D(Y)$ and temperature $\theta(Y)$ with central temperature $\theta_1 = 2.0$ showing a narrowed isothermal inner fluid component, in the same format as in Figure 12.}}
\end{figure}

\begin{figure}
\centerline{\includegraphics[width=120mm]{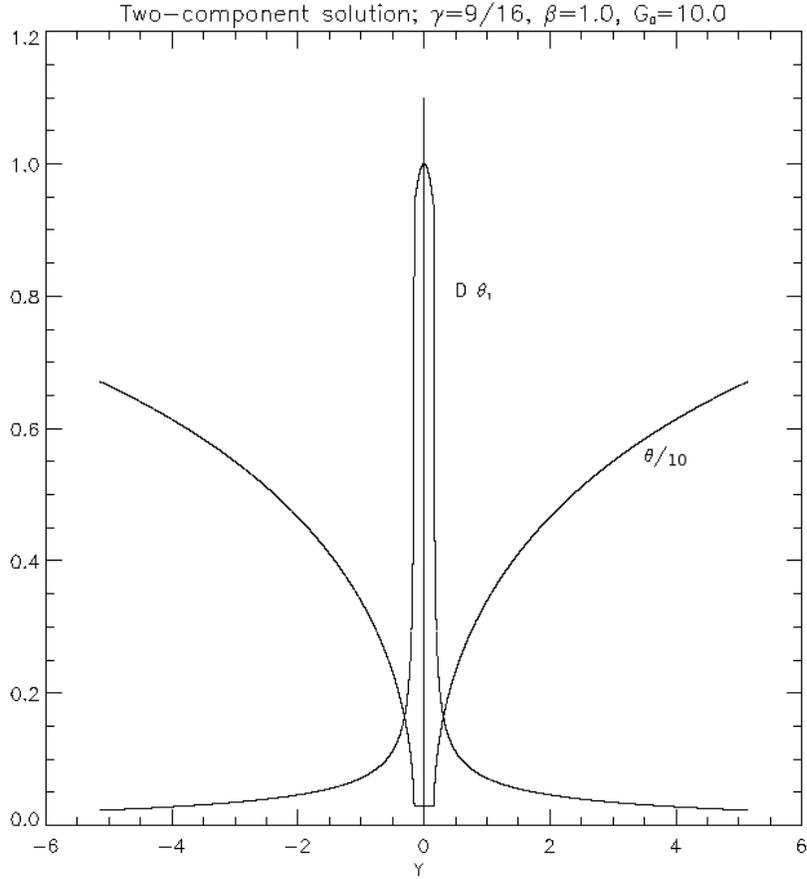}}
%\centerline{\includegraphics[width=130mm]{Fig14.eps}}
\caption{\small{The distributions of density $D(Y)$ and temperature $\theta(Y)$ with central temperature $\theta_1 = 0.3$ showing a greatly narrowed isothermal inner fluid component, in the same format as in Figure 12.   The temperature continues from its isothermal part into its $\gamma = 9/16$ with zero gradient, a feature not resolved at the scales of the figure.}}
\end{figure}

\end{document}